\documentclass[fleqn,usenatbib]{mnras}

\usepackage{newtxtext,newtxmath}

\usepackage[T1]{fontenc}

\DeclareRobustCommand{\VAN}[3]{#2}
\let\VANthebibliography\thebibliography
\def\thebibliography{\DeclareRobustCommand{\VAN}[3]{##3}\VANthebibliography}

\usepackage{graphicx}	
\usepackage{amsmath}	
\usepackage{siunitx}    
\usepackage{booktabs}   

\newcommand{\mach}{\mathcal{M}} 
\newcommand{\dd}{\mathrm{d}} 
\newcommand{\msun}{\,\mathrm{M_\odot}}
\newcommand{\nume}{\num[separate-uncertainty = true]}

\usepackage{scalerel,tikz}
\usetikzlibrary{svg.path}
\definecolor{orcidlogocol}{HTML}{A6CE39}
\tikzset{orcidlogo/.pic={
 \fill[orcidlogocol] svg{M256,128c0,70.7-57.3,128-128,128C57.3,256,0,198.7,0,128C0,57.3,57.3,0,128,0C198.7,0,256,57.3,256,128z};
 \fill[white] svg{M86.3,186.2H70.9V79.1h15.4v48.4V186.2z}
 svg{M108.9,79.1h41.6c39.6,0,57,28.3,57,53.6c0,27.5-21.5,53.6-56.8,53.6h-41.8V79.1z M124.3,172.4h24.5c34.9,0,42.9-26.5,42.9-39.7c0-21.5-13.7-39.7-43.7-39.7h-23.7V172.4z}
 svg{M88.7,56.8c0,5.5-4.5,10.1-10.1,10.1c-5.6,0-10.1-4.6-10.1-10.1c0-5.6,4.5-10.1,10.1-10.1C84.2,46.7,88.7,51.3,88.7,56.8z};
}}
\newcommand\orcidicon[1]{\href{https://orcid.org/#1}{\mbox{\scalerel*{
\begin{tikzpicture}[yscale=-1,transform shape]
\pic{orcidlogo};
\end{tikzpicture}
}{|}}}}

\usepackage{xcolor}
\usepackage[normalem]{ulem}

\DeclareSIUnit\parsec{pc}
\DeclareSIUnit\AU{AU}
\DeclareSIUnit\lightyear{ly}
\DeclareSIUnit\year{yr}

\title[Testing Turbulent Origin of IMF]{Testing the Turbulent Origin of the Stellar Initial Mass Function}

\author[D.~G.~Nam, C.~Federrath \& M.~R.~Krumholz]{
Donghee G.~Nam,$^{1}$\thanks{E-mail: u6836819@anu.edu.au}
Christoph Federrath$^{\orcidicon{0000-0002-0706-2306}\,1,2}$\thanks{E-mail: christoph.federrath@anu.edu.au} and
Mark R.~Krumholz$^{\orcidicon{0000-0003-3893-854X}\,1,2}$\thanks{E-mail: mark.krumholz@anu.edu.au}
\\
$^{1}$Research School of Astronomy and Astrophysics, Australian National University, Canberra, ACT~2611, Australia\\
$^{2}$ARC Centre of Excellence for Astronomy in Three Dimensions (ASTRO-3D), Canberra, ACT~2611, Australia
}

\date{Accepted XXX. Received YYY; in original form ZZZ}

\pubyear{2020}

\begin{document}
\label{firstpage}
\pagerange{\pageref{firstpage}--\pageref{lastpage}}
\maketitle

\begin{abstract}
Supersonic turbulence in the interstellar medium (ISM) is closely linked to the formation of stars, and hence many theories connect the stellar initial mass function (IMF) with the turbulent properties of molecular clouds. Here we test three turbulence-based IMF models (by Padoan \& Nordlund~2002, Hennebelle \& Chabrier~2008, and Hopkins~2012), which predict the relation between the high-mass slope ($\Gamma$) of the IMF, $\dd N/\dd \log M \propto M^{\Gamma}$ and the exponent $n$ of the velocity power spectrum of turbulence, $E_v(k)\propto k^{-n}$, where $n\approx2$ corresponds to typical ISM turbulence. Using hydrodynamic simulations, we drive turbulence with an unusual index of $n\approx 1$, measure $\Gamma$, and compare the results with $n\approx 2$. We find that reducing $n$ from 2 to 1 primarily changes the high-mass region of the IMF (beyond the median mass), where we measure high-mass slopes within the 95 per cent confidence interval of $-1.5<\Gamma<-1$ for $n\approx 1$ and $-3.7<\Gamma<-2.4$ for $n\approx 2$, respectively. Thus, we find that $n=1$ results in a significantly flatter high-mass slope of the IMF, with more massive stars formed than for $n\approx2$. We compare these simulations with the predictions of the three IMF theories. We find that while the Padoan \& Nordlund theory matches our simulations with fair accuracy, the other theories either fail to reproduce the main qualitative outcome of the simulations or require some modifications. We conclude that turbulence plays a key role in shaping the IMF, with a shallower turbulence power spectrum producing a shallower high-mass IMF, and hence more massive stars.

\end{abstract}




\section{Introduction} \label{sec:intro}

The stellar initial mass function (IMF), which describes the mass distribution of stars in a population at birth, plays a vital role in many fields of astrophysics. The literature generally agrees that the IMF has a power-law form $\dd N/\dd \log M \propto M^{\Gamma}$ in the high-mass end, with $\Gamma \approx -1.35$ \citep{Salpeter1955TheEvolution}, while there is an ongoing debate on the possible variations in observational estimates of the slope $\Gamma$ for extragalactic populations \citep{Bastian2010AVariations, Offner2014TheFunction, Krumholz2014TheFunction, Hopkins2018TheFunction}.

One popular candidate for determining the physics of the IMF is turbulence -- this is because the spectra of the molecular ISM, where stars are born, provide clear evidence for supersonic turbulent motions \citep{Larson1981TurbulenceClouds, Ossenkopf2002TurbulentClouds, Elmegreen2004InterstellarProcesses, Heyer2004TheClouds, Roman-Duval2011TheCalibration}. Thus, many theoretical models of the stellar IMF are based on the statistics of supersonic turbulence. \citet*{Padoan1997TheFunction} and \citet[hereafter PN02]{Padoan2002TheFragmentation} proposed that supersonic shocks create dense cores by sweeping through the ISM and compressing the gas. They then estimated the likelihood of the cores to be Jeans unstable and hence the mass distribution of collapsing cores, which may be closely linked to the IMF \citep{Andre2010FromSurvey, Offner2014TheFunction, Guszejnov2015MappingFunction}. \citet[hereafter HC08]{Hennebelle2008AnalyticalCores} and \citet[hereafter H12]{Hopkins2012TheDistribution} proposed derivations of the IMF using the \citet{Press1974FormationCondensation} and excursion set \citep{Bond1991ExcursionFluctuations} formalisms, respectively. In these models, one estimates the density variance as a function of size scale, and then determines the IMF by measuring the mass distribution of regions where the density is high enough for gravity to overcome various supporting mechanisms (such as thermal motions, turbulence, magnetic fields, and/or disc shear). The turbulence-regulated theories of the IMF by PN02, HC08, and H12 yield estimates for $\Gamma$ that are generally in good agreement with observed IMFs \citep{Miller1979TheNeighborhood,  Kroupa2001OnFunction, Chabrier2003GalacticFunction, Chabrier2005The2005, Kroupa2013ThePopulations, Offner2014TheFunction}, if the parameters are chosen carefully.

In these analytic models, the power-law index $n$ of the turbulent velocity power spectrum,\footnote{We define $E(k)$ to be the one-dimensional power spectrum, so that Kolmogorov turbulence corresponds to $n=5/3$.} $E_v(k) \propto k^{-n}$, appears as a critical factor that determines the high-mass power-law slope $\Gamma$. The narrow range of $n$ in nature ($5/3 \le n < 2$) \citep{Federrath2013OnTurbulence} can be used to argue for the relatively universal high-mass slope of the IMF produced by these models and seen in observations. However, the near universality of $n$ also makes it difficult to test any particular model's prediction for the relationship between $n$ and the IMF.  While the underlying functional relationship between $n$ and the IMF shape is fundamentally different in the different models, the small range of variation in $n$ yields a similarly small range in predicted IMFs.

Nonetheless, a handful of simulations have explored this question. \citet{Bate2009TheStructure} studied the effect of $n$ on the star formation within a collapsing molecular cloud by carrying out simulations with initial turbulent velocity fields characterised by $n=2$ and $n=4$, and concluded that the resultant IMFs show little dependence on $n$ overall. \citet*{Delgado-Donate2004TheFormation} conducted a set of similar simulations but in the context of low-mass ($5\msun$) core fragmentation, and also found that the initial choice of $n$ does not significantly affect the stellar IMF. \citet*{Goodwin2006StarSpectrum}, on the contrary, found that a shallower velocity power spectrum ($n$ closer to zero) leads to more fragmentation in their simulations of low-mass ($\sim 5\msun$) cores, although the statistical argument is weak due to the low number of sink particles used for the analysis ($N_\text{sink}<100$).
In the studies mentioned above, the authors varied only the initial velocity field, while the star formation commenced roughly after one free-fall time. The problem with this approach is that without continuous driving, most of the turbulent energy would dissipate away within a free-fall time \citep{Stone1998DissipationTurbulence, MacLow1998TheClouds, Elmegreen2004InterstellarProcesses, McKee2007TheoryFormation}, and $n$ would relax to the natural range of $5/3-2$. Therefore, while the choice of $n$ could affect the initial structure of the collapsing cloud, it would have little effect during the process of star formation. We conclude that the studies are insufficient for a direct comparison with the turbulent fragmentation theories.

The aim of this work is to test how well the turbulence-regulated IMF theories (PN02, HC08, and H12) predict the high-mass power-law slope of the IMF, by simulating star formation under hydrodynamic turbulence (i.e., without magnetic fields) with velocity power spectral index $n$ much different from what is observed in nature ($5/3-2$). We develop a turbulence driving module that is capable of driving and maintaining supersonic turbulence with arbitrary $n<2$, and create an artificial molecular cloud with $n=1$ in the computational domain. We measure the mass function of the stars, represented by sink particles, born under the $n=1$ turbulence, and compare it with the IMF from the typical $n\approx2$ supersonic turbulence. We assure the statistical significance of the study by collecting around 1000 stars represented by `sink particles' per setup through repeated simulations with different randomisation of the turbulence driving.

We note that the interaction between magnetic fields and the IMFs represents another point of difference that can be used to test the models. Magnetohydrodynamic (MHD) simulations show that magnetic fields have a variety of effects, including reducing the star formation rate and changing how gas fragments \citep{Padoan2014Infall-drivenProblem, Federrath2015InefficientFeedback, Haugbolle2018TheTurbulence,Krumholz2019TheFunction}. However, they are incorporated into IMF theories in differing ways. In the PN02 model, the presence of magnetic fields changes to which extent supersonic shocks compress the medium, which changes the mass spectrum of the density structures that may go on to collapse and form stars, whereas in the HC08 and H12 models the primary role of magnetic fields is to provide an additional form of pressure that makes it more difficult for structures to collapse. Although we present only hydrodynamic simulations here, in a forthcoming paper we explore the effects of magnetic fields as a complementary way of testing IMF theories.

The rest of the paper is organised as follows. We describe the simulation setup and the initial conditions in \S\ref{sec:methods}, and present the results in \S\ref{sec:result}. In \S\ref{sec:models} we compare our mass functions with the three turbulence-based IMF theories. We summarise our findings in \S\ref{sec:sum}.

\section{Numerical Methods} \label{sec:methods}

We simulate star formation within a turbulent, dense molecular cloud with the \textsc{flash4} adaptive mesh refinement (AMR) code \citep{Fryxell2000FLASHFlashes}. Here we use the HLL5R approximate Riemann solver \citep{Bouchut2010AWaves, Waagan2011ATests} and the multigrid Poisson gravity solver \citep{Ricker2008AMeshes} on a block-based PARAMESH AMR grid. We explain the turbulence driving method in \S\ref{sec:driving} and the sink particles in \S\ref{sec:sink}, then we outline the initial conditions and simulation procedure in \S\ref{sec:setup}.

\subsection{Turbulence driving} \label{sec:driving}

In order to drive turbulence with a prescribed velocity power spectrum of slope $n$, we add a time-varying acceleration field $\mathbf{F}_\text{stir}(\mathbf{x}, t)$ as a source term in the momentum equation \citep{Federrath2010ComparingForcing}. We utilise an Ornstein-Uhlenbeck process \citep{Eswaran1988AnTurbulence} to construct the driving field $\mathbf{F}_\text{stir}$ with an auto-correlation time matching the turbulent crossing time $T=L/2\sigma_v$, where $\sigma_v$ is the rms velocity dispersion. Inspired by observations \citep[e.g.][]{Ossenkopf2002TurbulentClouds,Elmegreen2004InterstellarProcesses,Brunt2009TurbulentClouds}, the usual procedure is to construct $\mathbf{F}_\text{stir}$ with only large-scale modes (i.e., to drive at wavenumbers\footnote{In this paper, $k$ is measured in units of the inverse box size, so $k=1$ corresponds to a mode with wavelength equal to the box scale $L$.} $k = \left|\mathbf{k}\right| \sim 2$) and let small-scale turbulence emerge naturally. The energy cascade in (supersonic) turbulence will distribute energy to smaller scales in such a way as to produce $n\approx 2$ \citep{Federrath2013OnTurbulence}.

Here, however, we want to construct velocity power spectra with $n$ significantly smaller than 2, in order to test theoretical predictions for the dependence of the IMF on $n$. Thus, we must inject energy on every resolvable scale, or in other words, the driving field needs to contain modes up to $k_N=L/(2\Delta x)$, where $\Delta x$ is the minimum computational cell size of the simulation. However, including all wavevectors within $2 \le k \le k_N$ is expensive since \textsc{flash} evaluates the acceleration field at each cell from the set of driving modes, and the number of modes in a wavenumber bin $[k,k+\dd k]$ is proportional to $k^2 \,\dd k$. To reduce the computational load, we take a heuristic approach by generating a stirring field that contains only a fraction of randomly-selected wavevectors, such that the number of modes between $k$ and $k+\dd k$ scales as $k^{0.5}\,\dd k$. This practice yields a significant gain in speed (by a factor of $\sim10^3$) while preserving the isotropy of $\mathbf{F}_\text{stir}$, and therefore the isotropy of the turbulence. The resultant driving field is constructed to have a natural mixture of solenoidal and compressive modes, which corresponds to the driving parameter $b\sim 0.4$ \citep{Federrath2010ComparingForcing}.

In order to run a set of simulations in which the power spectrum of the turbulent velocity field follows a power law with index $n=1$ or $n=2$, we construct the acceleration field $\mathbf{F}_\text{stir}$ with $\num{2.3e4}$ modes, randomly selected within $2\le k \le 256$. We show below that when the amplitude of each mode $A(\mathbf{k})$ is proportional to $k^{-0.9}$, the resulting turbulence power spectrum reaches a slope close to $n=1$. For the $n=2$ case, we use the same method, but with $A(\mathbf{k}) \propto k^{-2}$ to match the shape of the power spectrum of $\mathbf{F}_\text{stir}$ to that of the turbulent velocity typically observed in molecular clouds and simulations of supersonic turbulence \citep{Elmegreen2004InterstellarProcesses, McKee2007TheoryFormation, Federrath2013OnTurbulence}. Below we refer to simulations run with a driving field $A(\mathbf{k})\propto k^{-0.9}$ as N1 simulations, and those run with $A(\mathbf{k})\propto k^{-2}$ as N2 simulations. We show in Appendix~\ref{app:driverange} that the results we obtain for the N2 simulations using this driving procedure are nearly identical to those produced via the more common procedure of driving only at low $k$ \citep{Federrath2010ComparingForcing}, and allowing modes at higher $k$ to be produced by the turbulent cascade.

\subsection{Sink particles and AMR} \label{sec:sink}

In order to follow local collapse and accretion of gas, we use the sink particle method developed in \citet{Krumholz2004EmbeddingGrids} and extended by \citet{Federrath2010ModelingSPH}. \citet{Truelove1997TheHydrodynamics} showed that the local Jeans length $\lambda_J=(\pi c_s^2/G\rho)^{1/2}$, where $c_s$ is the sound speed, must be resolved with at least four grid cells to prevent artificial fragmentation of the collapsing gas. The sink particle technique ensures that the Jeans length is always sufficiently resolved on the highest level of AMR, and that only bound and collapsing gas is turned into sink particles. Gas above the sink particle density threshold
\begin{equation}
    \rho_\mathrm{sink} = \frac{\pi c_s^2}{G\lambda_J^2} = \frac{\pi c_s^2}{G r_\mathrm{sink}^2},
\end{equation}
with the sink particle radius $r_\mathrm{sink} = 2.5\Delta x_\text{min}$, is accreted, if the gas is bound and collapsing. Since not all overdense regions that satisfy the above density condition will collapse, we adopt an additional set of sink creation criteria from \citet{Federrath2010ModelingSPH} to avoid artificial sink particle formation.

For dense regions that are not yet on the highest level of AMR, we refine based on the local Jeans density, to better resolve the gravitational collapse. In our simulations, $\lambda_J$ is resolved with at least 16 cells in all dimensions, in order to capture some solenoidal motions of the turbulence inside the Jeans scale \citep{Federrath2011ATurbulence}.

\subsection{Simulation setup} \label{sec:setup}

We simulate a small section of a molecular cloud within a three-dimensional periodic computational domain of length $L=\SI{2}{\parsec}$, mean gas density $\rho_0=\SI{1.31e-20}{\g\per\cubic\cm}$, and thus the cloud mass $M_\text{cloud}=\rho_0L^3=1550\msun$. The base-grid resolution is $N_\text{base}=512^3$ grid cells, with two additional levels of AMR, which leads to a maximum effective resolution of $2048^3$ cells, i.e., a minimum cell size of $\Delta x\approx 200$~AU. At this resolution we cannot capture detailed small-scale structures and physics such as protostellar discs and radiative feedback. While radiative feedback may be crucial for setting the characteristic mass of the IMF \citep[but see \citealt{Haugbolle2018TheTurbulence}]{Bate2009TheStructure, Offner2009TheFormation, Krumholz2011OnMasses,Krumholz2016WhatSimulations, Federrath2017ConvergingStars}, at least in the theoretical models that we aim to test it has little effect on the high-mass slope of the IMF. We therefore focus solely on determining the role of the turbulence power spectrum for the high-mass tail of the IMF, and compare to predictions from IMF theories. We assume isothermal gas with constant global sound speed $c_s=\SI{0.2}{\km\per\second}$, and drive the turbulence to an rms Mach number $\mach=\sigma_v/c_s=5$ for all simulations. This ensures that all simulations have identical total kinetic energy, and thus the same global virial parameter \citep{Bertoldi1992Pressure-confinedClouds}, $\alpha_\text{vir} = 5 \sigma_v^2 L / (6GM)=0.25$, and free-fall time $t_\text{ff} = \sqrt{3\pi/(2G\rho_0)} = \SI{0.58}{\mega\year} = 0.594\,T$. While our choice of mean density is a factor of 2--3 higher than the \citet{Larson1981TurbulenceClouds} relation\footnote{According to the Larson relation, a cloud with $L=\SI{2}{\parsec}$ has $n(\text{H}_2)=\SI{1600}{\per\cm\cubed}$, or $\rho_0=\SI{5.4e-21}{\g\per\cm\cubed}$; however, there is substantial scatter around this relation \citep{Larson1981TurbulenceClouds, Falgarone1992TheClouds}.}, the choice of scaling cannot affect the shape of the IMF, which is the quantity of interest for us. We also emphasise that this commonly used approximation of $\alpha_\text{vir}$ is based on the uniform spherical approximation, and the geometry of our simulations is much different from a sphere of gas. The calculated value of $\alpha_\text{vir}$ based on its definition, $2E_\text{kin}/|E_\text{grav}|$, is more than an order of magnitude higher than the approximated value of 0.25, and is dependent on turbulence parameters such as $b$ and $n$ \citep{Federrath2012TheObservations}. This discrepancy is particularly strong for the simulations with $n=1$, which, as we show below, develop significantly less large-scale density structure than the $n=2$ case, and thus have weaker self-gravity than one might otherwise expect.

All simulations begin with uniform density distribution $\rho(\mathbf{x})=\rho_0$ and zero velocity $\mathbf{v}(\mathbf{x})=0$. We let the supersonic turbulence grow by running the models without self-gravity for two turbulent crossing times $2\,T$ \citep{Federrath2012TheObservations}, after which, gravity is turned on and sink particles are allowed to form in bound, collapsing regions of the cloud. We aim to collect around 1000 sink particles for each case to obtain tight statistical constraints on the slopes of the mass functions of the sink particles. For this reason, we run fourteen simulations where we drive with a field $A(\mathbf{k})\propto k^{-0.9}$ in order to produce $n\approx 1$ (N1A--N1N) and eight simulations where we drive with $A(\mathbf{k})\propto k^{-2}$ and thus produce $n\approx 2$ (N2A--N2H). Table~\ref{tab:params} summarises the key input parameters and derived quantities.

\begin{table*}
    \centering
    \caption{Key simulation parameters and measured quantities.}
    \begin{tabular}{rrrrrrrrr}
    \toprule
    ID & $n$ & $\mach$ & $N_\text{sink}$ & $\text{SFR}_\text{ff}$ & $m_{50}$ & $m_{84}$ & $m_{98}$ \\
    (1) & (2) & (3) & (4) & (5) & (6) & (7) & (8) \\
    \midrule
    \multicolumn{1}{l}{\textbf{N1}} & & & & & \\
    \phantom{\textbf{N1}}A & \nume{0.93(3)} & 5.0 & 102 & 0.14 & \num{4.7e-04} & \num{1.7e-03} & \num{5.6e-03} \\
    B & \nume{0.91(3)} & 4.8 & 99 & 0.53 & \num{5.3e-04} & \num{1.5e-03} & \num{6.5e-03} \\
    C & \nume{0.92(2)} & 4.9 & 67 & 0.21 & \num{9.1e-04} & \num{2.4e-03} & \num{7.1e-03} \\
    D & \nume{0.96(2)} & 4.9 & 57 & 0.14 & \num{4.3e-04} & \num{5.0e-03} & \num{1.0e-02} \\
    E & \nume{0.95(2)} & 5.0 & 74 & 0.17 & \num{3.5e-04} & \num{1.8e-03} & \num{1.1e-02} \\
    F & \nume{0.97(2)} & 5.0 & 97 & 0.22 & \num{3.9e-04} & \num{2.0e-03} & \num{7.1e-03} \\
    G & \nume{0.97(3)} & 5.0 & 56 & 0.14 & \num{4.6e-04} & \num{2.6e-03} & \num{1.5e-02} \\
    H & \nume{0.92(3)} & 4.9 & 64 & 0.13 & \num{6.3e-04} & \num{3.7e-03} & \num{6.2e-03} \\
    I & \nume{0.98(2)} & 4.9 & 55 & 0.04 & \num{4.5e-04} & \num{2.5e-03} & \num{1.4e-02} \\
    J & \nume{0.98(3)} & 5.0 & 56 & 0.31 & \num{9.9e-04} & \num{4.1e-03} & \num{6.8e-03} \\
    K & \nume{0.98(3)} & 5.0 & 66 & 0.13 & \num{4.6e-04} & \num{3.1e-03} & \num{9.1e-03} \\
    L & \nume{0.86(3)} & 4.8 & 76 & 0.17 &  \num{4.7e-04} & \num{2.1e-03} & \num{1.1e-02} \\
    M & \nume{0.92(3)} & 5.0 & 54 & 0.19 & \num{4.5e-04} & \num{3.1e-03} & \num{1.6e-02} \\
    N & \nume{0.95(2)} & 5.0 & 64 & 0.20 & \num{4.3e-04} & \num{3.8e-03} & \num{7.4e-03} \\
    \midrule
    \textbf{total} & \textbf{\nume{0.95(1)}} & \textbf{\nume{4.9(1)}} & \textbf{987} &  & \textbf{\num{4.9e-4}} & \textbf{\num{2.5e-3}} & \textbf{\num{1.0e-2}} \\
    \midrule
    \multicolumn{1}{l}{\textbf{N2}} \\
    \phantom{\textbf{N2}}A & \nume{1.80(1)} & 5.2 & 114 & 0.27 & \num{5.0e-04} & \num{1.9e-03} & \num{4.3e-03} \\
    B & \nume{1.87(1)} & 4.5 & 110 & 0.30 & \num{6.3e-04} & \num{1.8e-03} & \num{3.3e-03} \\
    C & \nume{1.89(1)} & 4.9 & 137 & 0.40 & \num{4.7e-04} & \num{1.3e-03} & \num{3.4e-03} \\
    D & \nume{1.89(1)} & 4.7 & 112 & 0.30 & \num{6.3e-04} & \num{1.6e-03} & \num{3.1e-03} \\
    E & \nume{1.91(1)} & 4.7 & 126 & 0.34 & \num{4.7e-04} & \num{1.4e-03} & \num{3.6e-03} \\
    F & \nume{1.84(1)} & 5.0 & 113 & 0.36 & \num{3.8e-04} & \num{1.6e-03} & \num{5.7e-03} \\
    G & \nume{1.87(1)} & 4.8 & 109 & 0.39 & \num{5.0e-04} & \num{1.9e-03} & \num{3.1e-03} \\
    H & \nume{1.85(1)} & 4.7 & 105 & 0.32 & \num{5.7e-04} & \num{1.7e-03} & \num{4.4e-03} \\
    \midrule 
    \textbf{total} & \textbf{\nume{1.86(1)}} & \textbf{\nume{4.8(2)}} & \textbf{926} & & \textbf{\num{5.0e-4}} & \textbf{\num{1.7e-3}} & \textbf{\num{4.0e-3}} \\
    \bottomrule
    \end{tabular} \vspace{0.1cm}\\
    {\raggedright \emph{Notes.} (1) simulation name; (2--3) power-law index $n$ and rms Mach number $\mach$ measured after two turbulent crossing times; (4--5) the number of sink particles and star formation rate (SFR) per free-fall time recorded at the star formation efficiency (SFE) of 10 per cent; (6--8) 50th, 84th, and 98th percentiles of the SMF, where masses are measured as $m = M_{\rm sink}/M_{\rm cloud}$.
    \par}
    \label{tab:params}
\end{table*}

\section{Results} \label{sec:result}

In this section we analyse the results of the simulations summarised in Table~\ref{tab:params}. First we examine the statistics of the velocity and density fields in \S\ref{sec:stats}, and verify that our turbulence driving method produces a range of power-law slopes as desired. We then study how the modified turbulence affects molecular cloud morphology in \S\ref{sec:cloud}. We discuss the star formation rate and temporal evolution of the simulations in \S\ref{sec:sfr}, and finally, we construct the sink mass function (SMF) and calculate its power-law slope $\Gamma$ in \S\ref{sec:imf}. Although we carry out simulations in physical units, as described in \S\ref{sec:setup}, we note that, since they are isothermal, the simulations themselves are dimensionless and can be re-scaled to arbitrary length and mass scales. For this reason, in this section we will report all results in dimensionless units, i.e., we will report all masses as fractions of $M_{\rm cloud}$, all lengths as fractions of $L$, and so forth, since these ratios are independent of the choice of dimensional scaling.

\subsection{Velocity and density statistics} \label{sec:stats}

\begin{figure}
    \centering
    \includegraphics[width=\columnwidth]{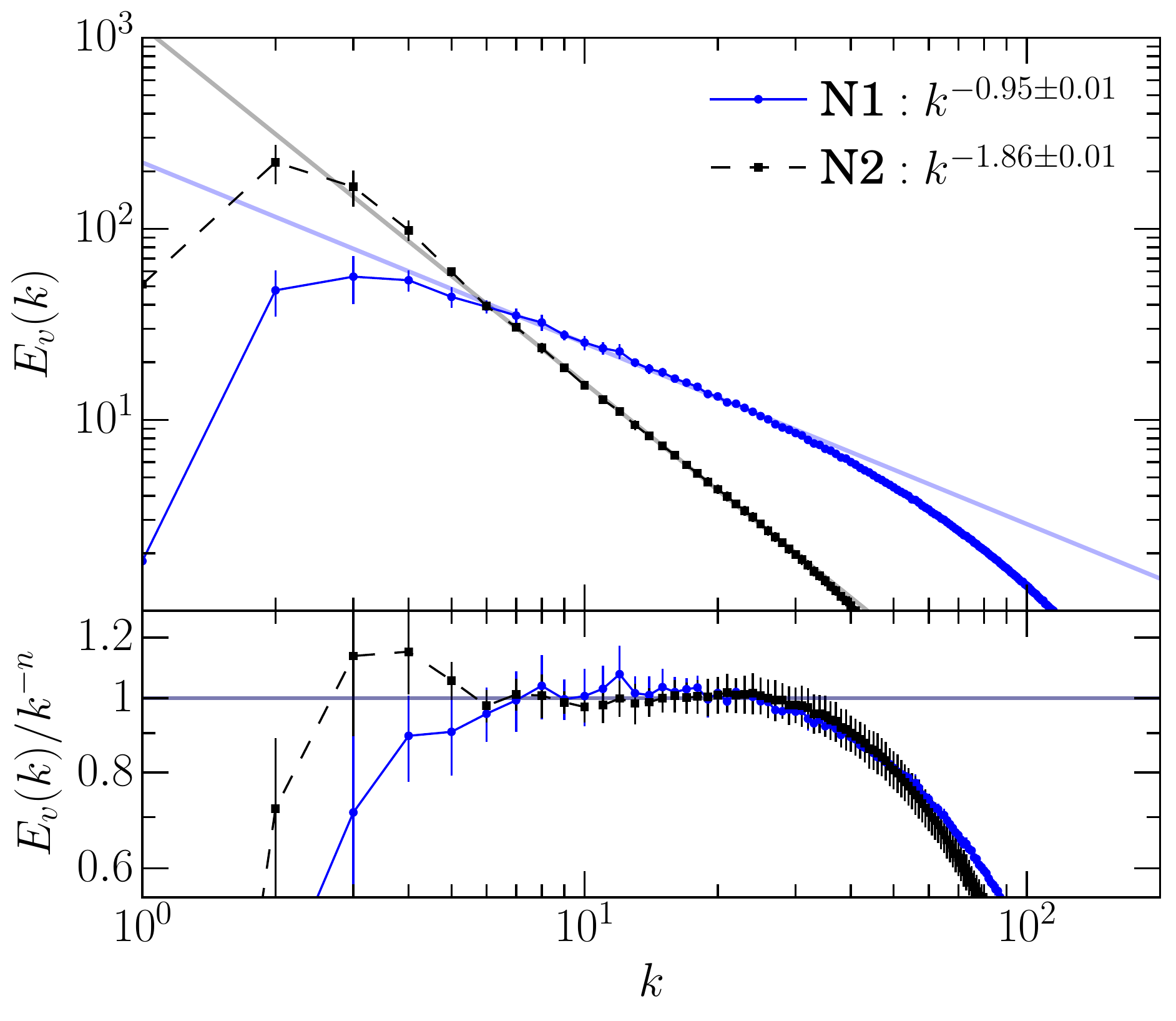}
    \caption{Turbulent velocity power spectra $E_v(k)$ (top) and the compensated power spectra $E_v(k)/k^{-n}$ (bottom), for N1 ($n=1$; blue solid line) and N2 ($n=1.9$; black dashed line). The vertical lines indicate the $1\sigma$ range of variation within the simulations, and the thick transparent lines in the top panel are power-law fits over the range $5\le k\le 30$. The $y$-axes in both panels have arbitrary units, and the compensated power spectra are normalised so that their means within the fitting range are both equal to 1.}
    \label{fig:ps}
\end{figure}

\begin{figure}
    \centering
    \includegraphics[width=\columnwidth]{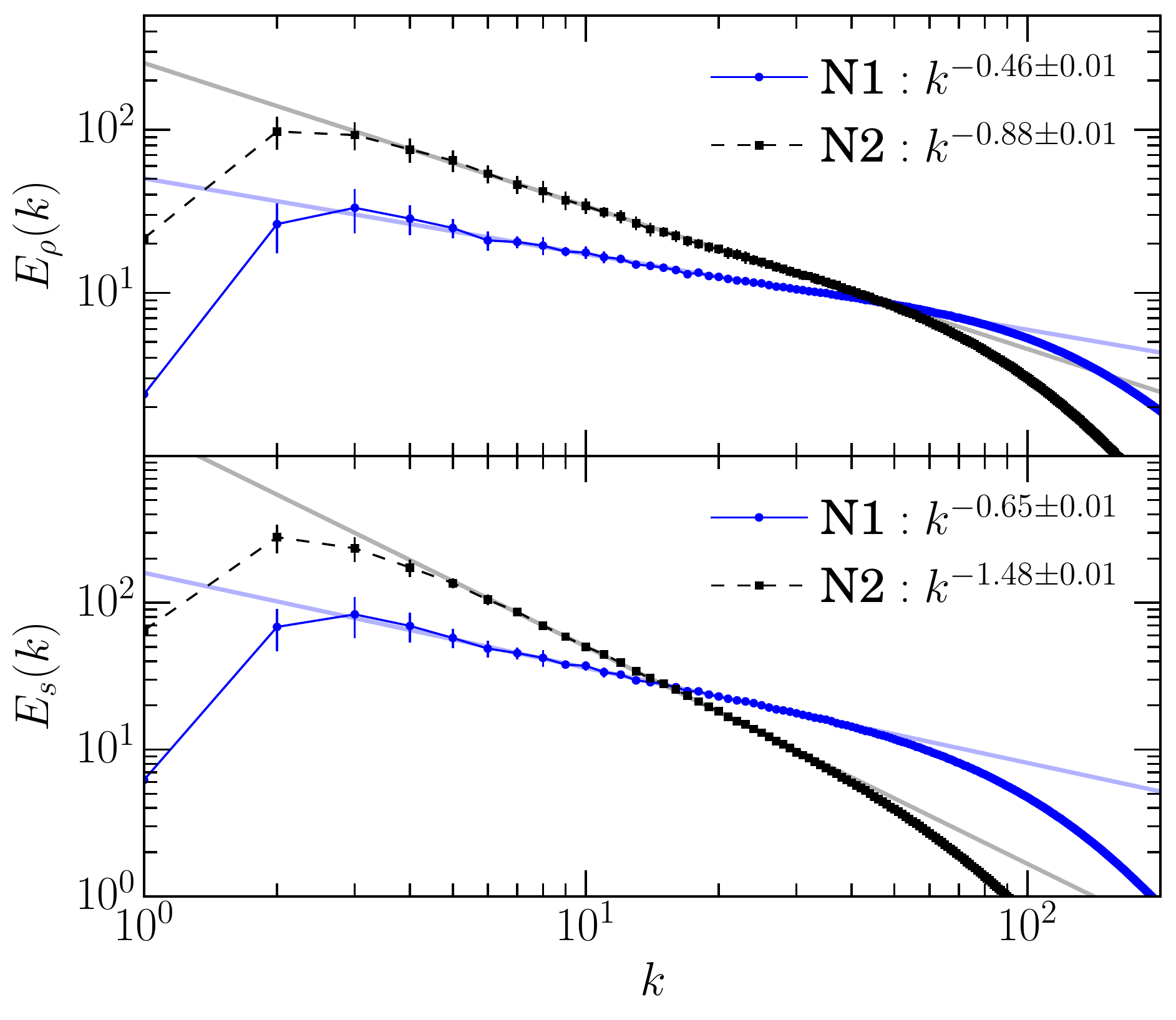}
    \caption{Power spectra of the density $\rho$ (top) and logarithmic density $s=\ln(\rho/\rho_0)$ (bottom) for N1 and N2 runs. Symbols and fitting methods are identical to those used in Fig.~\ref{fig:ps}. The $y$-axes have arbitrary units.}
    \label{fig:psdens}
\end{figure}

To confirm that the simulations reach the intended values of the velocity power spectral index $n$ we measure the velocity power spectra $E_v(k)$ of the simulations at $t=2\,T$, i.e., when the turbulence would be fully developed and gravitational collapse begins. We interpolate the AMR grid to a $512^3$ uniform grid (i.e., at the base-grid resolution) when calculating the power spectra. Fig.~\ref{fig:ps} shows the resulting power spectra, averaged over each set of runs, i.e., the line labelled N1 in the plot is the average power spectrum of runs N1A--N1N, and similarly for N2. For both sets of simulations, the power spectra show a power-law dependence on $k$ over a broad range of length scales until $k \sim 30$, beyond which numerical dissipation begins to take effect. We therefore estimate the slope of the power-law by fitting the velocity power spectrum $E_v(k)$ over the range $5\le k \le 30$. We find best-fit values $E_v(k)\propto k^{\nume{-0.95(1)}}$ for N1 and $k^{\nume{-1.86(1)}}$ for N2, as shown in the top panel of Fig.~\ref{fig:ps}. The value of $n$ for N1 is in good agreement with our target, while the one for N2 is slightly shallower, because of the low target Mach number \citep[see e.g.][for comparison]{Kritsuk2007TheTurbulence,Federrath2010ComparingForcing}. Nonetheless, it is clearly steeper than the result for N1. We also present the compensated power spectra, in the bottom panel of Fig.~\ref{fig:ps}, to better visualise the deviations from the power-law scaling. In both simulations, $E_v(k)$ follows the scaling law very well within the fitting range. We conclude that we successfully drive and maintain turbulence such that its velocity power spectrum is a power-law with an index of $-1$ or $\approx -2$ for a broad range of length scales, as required for the experiment we wish to perform.

In the top panel of Fig.~\ref{fig:psdens} we plot the density power spectra, $E_\rho(k)$, which we measure and fit exactly as we do the velocity field, for N1 and N2 runs. We find turbulence with $n=1$ has considerably less power on large spatial scales (small $k$) than with $n=1.9$, due to the weaker large-scale turbulence. More interestingly, the total variance of the density fluctuations,
\begin{align}
    \left<\rho^2 \right> = \int E_\rho(k) \, \dd k,
\end{align}
for N1 simulations is about 20 per cent lower than for the N2 counterpart, despite the fact that the total velocity fluctuation $\sigma_v^2 = (\mach c_s)^2$ is equal in both cases.

The bottom panel of Fig.~\ref{fig:psdens} shows the power spectra of the logarithmic density $s=\ln(\rho/\rho_0)$, $E_s(k)$, for N1 and N2 runs. We find the spectral index of $E_s(k)$, which we denote as $-n'$, to be $n'=\nume{0.65(1)}$ for N1 and $n'=\nume{1.48(1)}$ for N2. Although the exact scaling exponent of the density power spectrum remains in debate (our result for $E_\rho(k)$ is similar to that of \citet{Kim2005DensityFlows} and slightly shallower than found by \citet{Konstandin2016MachField}), it is important to note that $n'$ does not equal $n$ for both simulations. This contradicts a core assumption in the HC08 model and we discuss the impact this has on the shape of the HC08 IMF in detail in \S\ref{sec:hc08}.

\subsection{Cloud structure} \label{sec:cloud}

\begin{figure*}
    \centering
    \includegraphics[width=\linewidth]{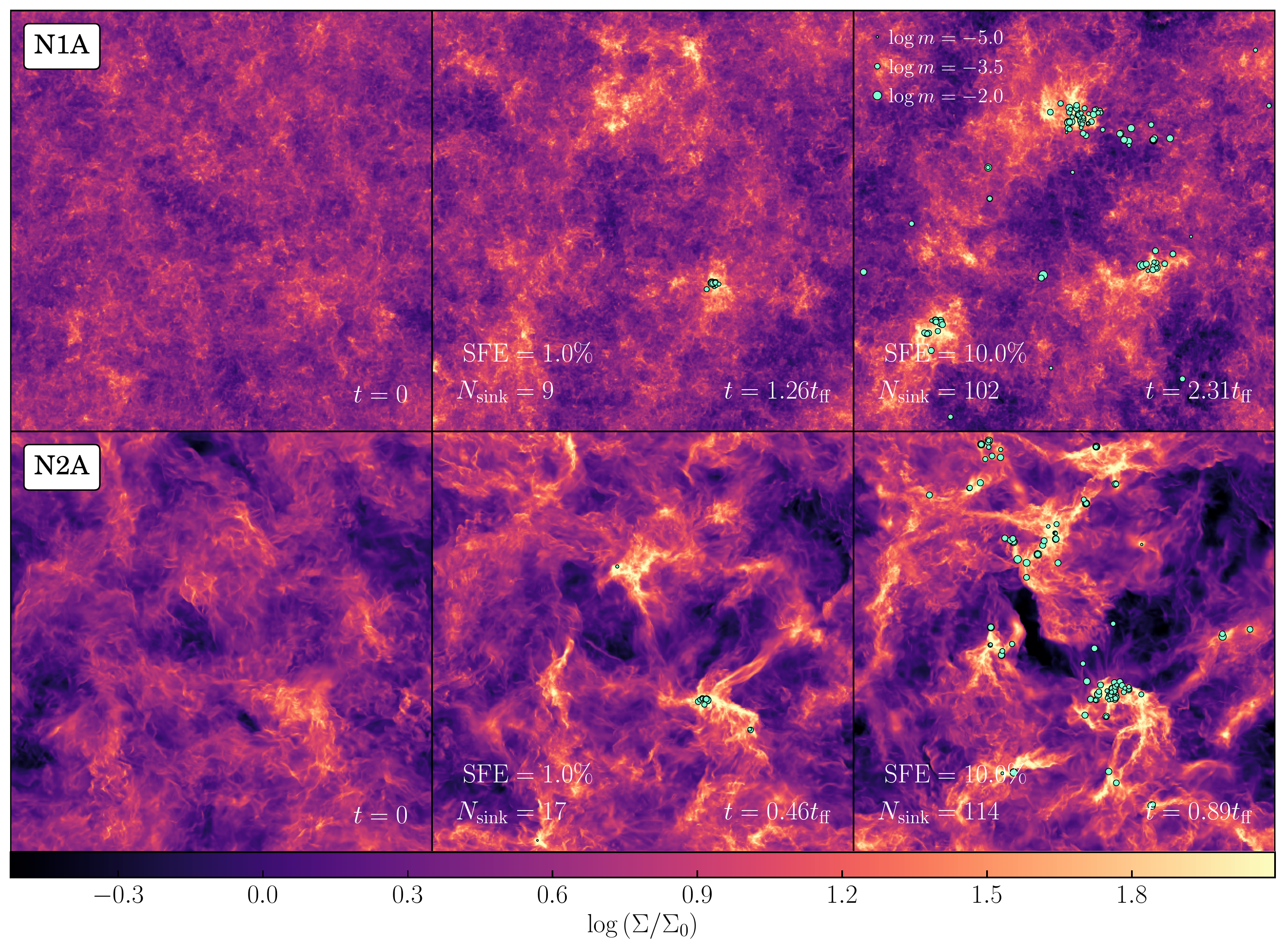}
    \caption{Column density maps extracted from N1A (top) and N2A (bottom), at the point when we turn on self-gravity (left) and at the times when the simulations reach a SFE of 1\% (middle) and 10\% (right). The colour scale is logarithmic and ranges from $\Sigma=0.3\Sigma_0$ (black) to $125\Sigma_0$ (white), where $\Sigma_0 = \rho_0 L$. We plot sink particles as cyan circles on top of the density projections, with sizes proportional to the logarithm of their mass $m=M_\text{sink}/M_\text{cloud}$ as indicated in the legend. Panels are annotated with the number $N_{\rm sink}$ of sink particles present in the frame and time $t$ of the simulation, where $t=0$ corresponds to the time at which we turn on self-gravity.}
    \label{fig:proj}
\end{figure*}

Fig.~\ref{fig:proj} compares the column density distributions of run N1A (top) with N2A (bottom). The left panels show the structure at time $2\,T$, immediately before we turn on self-gravity. This figure confirms our speculations based on Fig.~\ref{fig:psdens}: there exist large ($k\sim 5$) density structures in the cloud with $n\approx 2$, but such structures are much less prominent in the $n=1$ model. Instead, small-scale velocity perturbations dominate the cloud, which prevent large-scale density structures from forming. As a result the overall level of density perturbation in N1A is smaller than in N2A, which explains why the integral of $E_\rho(k)$ is lower for $n=1$.

The dominance of small-scale turbulence in N1A continues after the self-gravity is switched on, as shown in the middle and right panels of Fig.~\ref{fig:proj}. While the standard supersonic turbulence ($n\approx 2$) allows gas to collapse into dense filaments, inside which dense protostellar cores emerge, gas in the $n=1$ turbulence collapses in a fairly different manner. We no longer observe gas filaments, but dense, quasi-spherical patches of gas, and fragmentation happens inside these patches. There are two explanations for the lack of gas filaments: run N1A lacks low-$k$ supersonic shocks that compresses gas in one dimension over large spatial scales, and the excessive amount of turbulent energy in high-$k$ modes would quickly destroy the filaments.

\subsection{Star formation rate} \label{sec:sfr}

\begin{figure}
    \centering
    \includegraphics[width=\columnwidth]{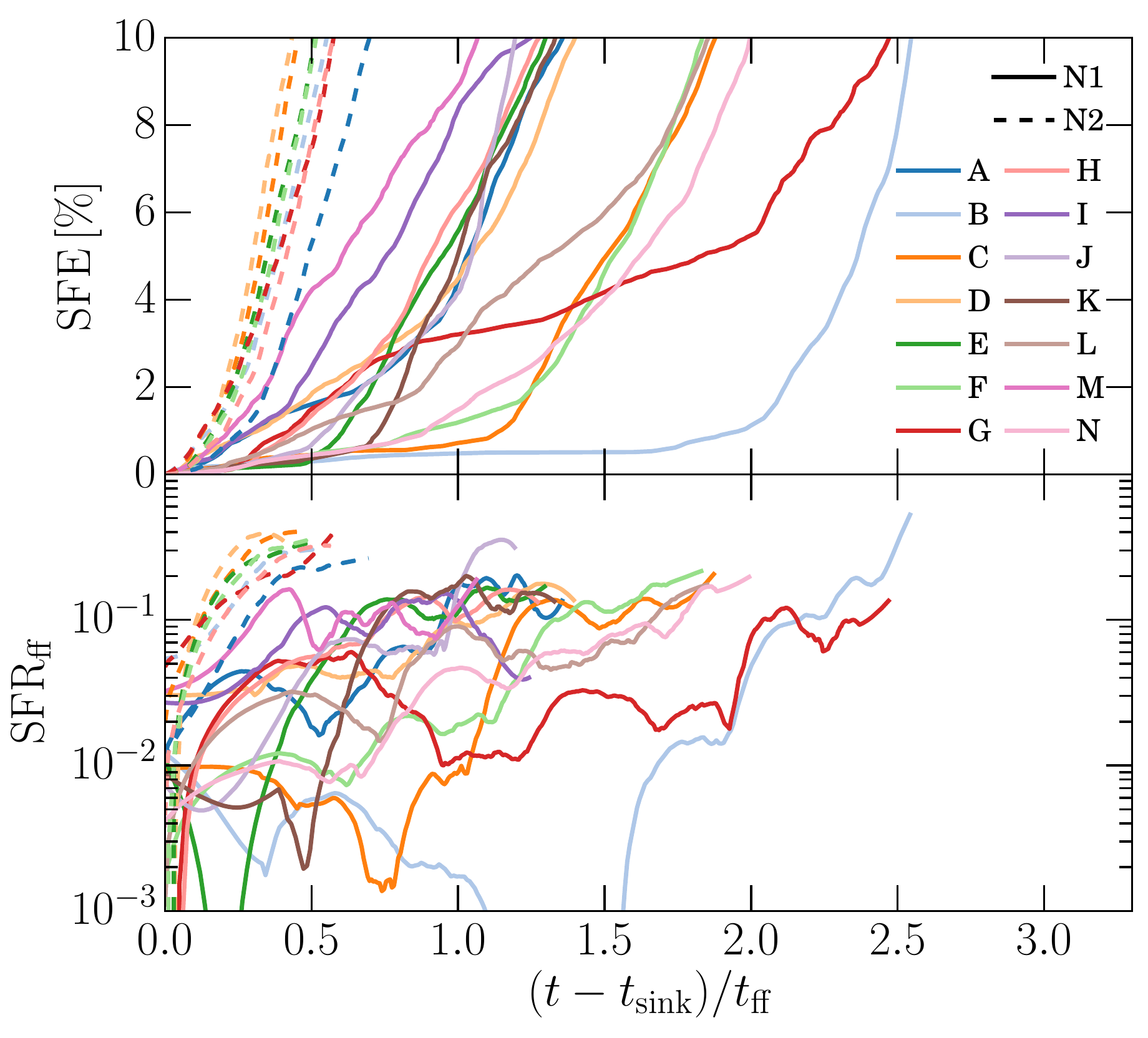}
    \caption{Star formation efficiency (top) and star formation rate measured per free-fall time (bottom) plotted as a function of time since the formation of the first sink particle, which we denote $t_{\rm sink}$. Solid lines are simulations with $n=1.9$ and dashed lines are for $n=1$. In general, N1 simulations evolve more slowly than N2 simulations and have lower star formation rates.}
    \label{fig:sfr}
\end{figure}

We note in Fig.~\ref{fig:proj} that star formation is much slower in turbulence with $n=1$. N2A arrives at a star formation efficiency ($\text{SFE}=M_\text{sink}/M_\text{cloud}$) of 10 per cent after $0.89\,t_\text{ff}$, whereas it takes $2.31\,t_\text{ff}$ for N1A to convert the same amount of mass into sinks. In order to show that this is a general result and not just the case for N1A versus N2A, we plot the temporal evolution of the SFE and the star formation rate (SFR) measured per free-fall time $\text{SFR}_\text{ff} = \dd\,\text{SFE}/\dd(t/t_\text{ff})$ for all our simulations in Fig.~\ref{fig:sfr}. We observe that it takes an average of approximately 0.5 free-fall times for the N2 simulations to go from the formation of their first sink particle to the time when the SFE reaches 10\% and we stop the simulation, whereas this number grows to $\sim 1.7\,t_\text{ff}$ for N1 simulations. Similarly, we see that turbulence with $n=1$ keeps $\text{SFR}_\text{ff}\lesssim 0.2$ throughout most of the simulations, while for the N2 simulations with $n=1.9$ we have $\text{SFR}_\text{ff}\sim 0.3$.

One distinct and noteworthy feature is that some N1 simulations show a longer period of near-quiescence, even after the first sink particle appears, before the onset of vigorous star formation. Simulation N1B (light blue solid line in Fig.~\ref{fig:sfr}) is the most extreme example of this: even after the first sink forms, this run remains at $\text{SFE}\approx 0.5\%$ for almost 2 free-fall times, but then the $\text{SFR}_\text{ff}$ peaks at 0.54 near the end of the run. On the contrary, all N2 simulations show a much more regular pattern where star formation begins slowly, but then $\text{SFR}_{\text{ff}}$ rapidly increases over $\lesssim 1$ free-fall time.

\subsection{Mass function of the sink particles} \label{sec:imf}

\begin{figure}
    \centering
    \includegraphics[width=\columnwidth]{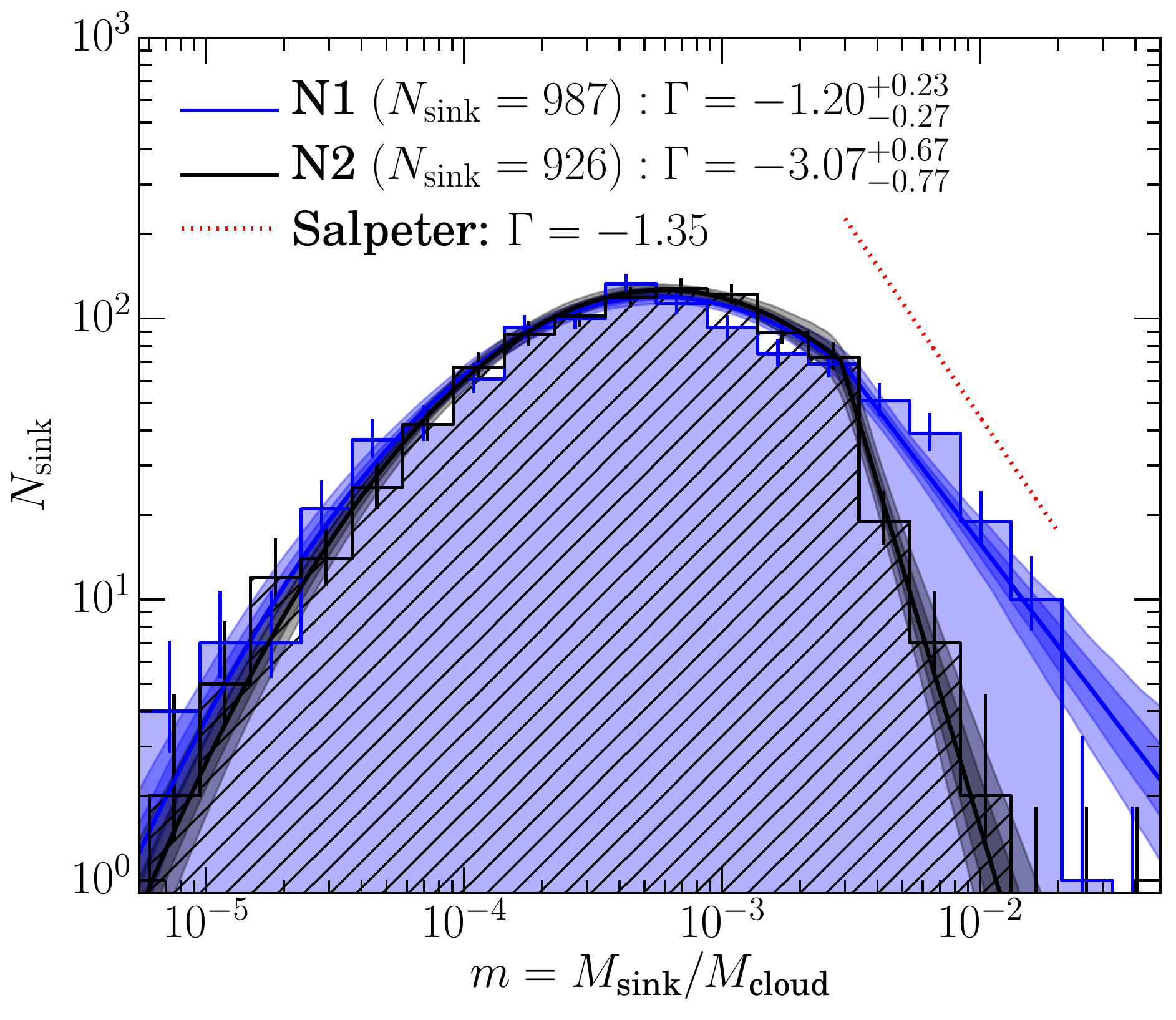}
    \caption{Logarithmic mass function $\dd N/ \dd\log{m}$ ($m=M_\text{sink}/M_\text{cloud}$ is the sink mass relative to the cloud mass) of the sink particles from the N1 (blue shaded histogram) and N2 (black hatched histogram) simulations, at $\text{SFE}=10\%$. The error bars on the histograms indicate the 68\% confidence interval for each bin. The solid lines show the median values of the posterior PDF obtained from the MCMC fitting, with the surrounding shaded regions representing the 68\% (thick shades) and 95\% (light shades) confidence intervals determined from the MCMC fit. We also report the median values for the high-mass power-law slope $\Gamma$, with the 2nd to 98th percentile ranges in the legend. The red dotted line corresponds to the \citet{Salpeter1955TheEvolution} slope ($\Gamma=-1.35$). We find that the power-law slope of the SMFs generated from the simulations are shallower (for N1) and significantly steeper (for N2) than the Salpeter slope. Thus, the turbulence power spectrum plays a key role in controlling the high-mass slope of the IMF.}
    \label{fig:imf}
\end{figure}

\begin{figure}
    \centering
    \includegraphics[width=\columnwidth]{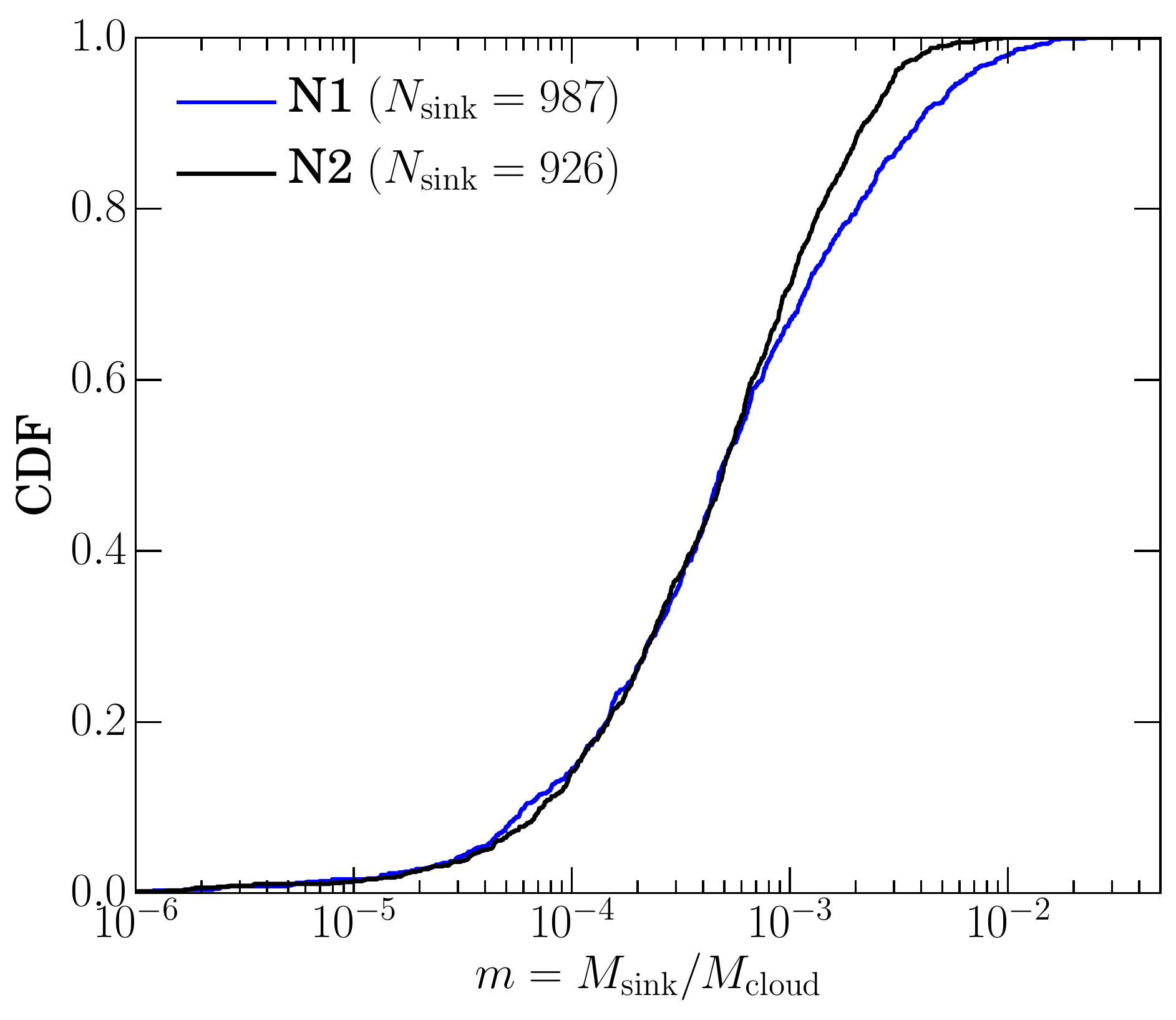}
    \caption{Cumulative distribution function (CDF) of the sink masses for N1 (blue) and N2 (black) simulations. We observe that the CDFs with $n=1$ and 1.9 disagree only within the high-mass (beyond the median mass) region. N1 simulations produce a significantly more top-heavy CDF.}
    \label{fig:cdf}
\end{figure}

We collect sink particles from the simulations when they reach $\text{SFE}=10\%$ and construct the Sink Mass Functions (SMFs) $\dd N/\dd \log m$ for each value of $n$, where $m$ is the relative mass $m=M_\text{sink}/M_\text{cloud}$ of the sinks. Fig.~\ref{fig:imf} shows the resultant SMFs, which span three orders of magnitude in mass and thus provide a sufficient dynamic range to identify differences between the N1 and N2 cases at high confidence. Quantitatively, we form sinks as small as $m= \num{5e-7}$ ($M_\mathrm{sink}= \num{8e-4}\msun$), and as large as $m= \num{2.5e-2}$ ($M_\mathrm{sink}= 40\msun$); the lower cutoff is imposed by the resolution of the simulation, while the upper one is due to the finite amount of mass contained in the periodic box. We observe that the N1 simulations generate significantly more sinks with $m\gtrsim\num{5e-3}$ than the N2 simulations. This makes the high-mass fall-off in N1 slightly shallower than that of the \citet{Salpeter1955TheEvolution} IMF, while the N2 SMF shows high-mass scaling visibly steeper than the Salpeter slope. The characteristic mass where the IMF peaks ($m\approx 10^{-3}$), on the other hand, appears to be fairly insensitive to the velocity power spectral index. 

We compare the cumulative mass functions for the N1 and N2 runs in Fig.~\ref{fig:cdf}. The figure clearly shows that the mass distributions are statistically indistinguishable below the median mass, but that the cumulative SMF for N1 is skewed significantly towards higher mass compared to that for N2. To demonstrate this quantitatively, we report the values of the 50th, 84th, and 98th percentile of the SMF in Table~\ref{tab:params}. While we find that the median masses are almost identical for N1 and N2 ($m_{50}=\num{4.9e-4}$ for N1 and $\num{5.0e-4}$ for N2), the 86th and 98th percentile masses widely differ, as one can find from Table~\ref{tab:params}. We also conduct a Kolmogorov-Smirnov (KS) test comparing the SMFs. If we compare only the parts of the distribution below the median mass, the test returns a $p$-value $p=0.59$, consistent with the hypothesis that the N1 and N2 data are drawn from the same parent distribution. However, if we instead compare the full SMFs, we obtain $p\sim 10^{-8}$. These statistics provide additional evidence for our speculation that altering the turbulence spectral index primarily affects the high-mass tail of the IMF.

Finally, in order to quantitatively measure the difference in the slope of the SMFs ($\Gamma$), we use the Markov Chain Monte-Carlo (MCMC) sampler \texttt{emcee} \citep{Foreman-Mackey2013EmceeHammer} to fit the SMFs to a \citet{Chabrier2005The2005}-like functional form for the IMF,
\begin{align}
    \frac{\dd N}{\dd \log m} = A_1
    \begin{cases}
        \dfrac{1}{\sqrt{2\pi \sigma^2}} \exp\left[-\dfrac{\left(\log m - \log m_0\right)^2}{2\sigma^2} \right], &\quad m<m_T, \\
        A_2 m^{\Gamma}, &\quad m\ge m_T,
    \end{cases} \label{eq:fit}
\end{align}
with four free parameters $\boldsymbol\theta = (m_0, \sigma, m_T, \Gamma)$, where $m_0$ and $\sigma$ are respectively the peak and standard deviation of the log-normal part, $m_T$ is the transition point between the log-normal and power-law part, and $\Gamma$ is the power-law slope. $A_1$ is a normalisation constant, set by the total mass in stars, and $A_2$ is set so as to ensure continuity at $m_T$.\footnote{We note that the derivative of Eq.~(\ref{eq:fit}) is not necessarily continuous at $m=m_T$. We allow this possibility to ensure that the slope we find for the power-law portion of the IMF at high masses is not forced to some particular value by a requirement that it match the slope favoured by the sub-peak sink population, which dominates the total number of sink particles, and thus the likelihood function.} The posterior probability distribution for $\boldsymbol\theta$ is given by Bayes' Theorem,
\begin{align}
    P(\boldsymbol\theta|\{m_\text{sink}\}) = \frac{P(\boldsymbol\theta)P(\{m_\text{sink}\}|\boldsymbol\theta)}{\int P(\boldsymbol\theta')P(\{m_\text{sink}\}|\boldsymbol\theta') \,\dd \boldsymbol\theta'},
\end{align}
where the likelihood function for a given set of parameters $\boldsymbol\theta$ and sink masses $\{m_\text{sink}\}$ is
\begin{align}
    P(\{m_\text{sink}\}|\boldsymbol\theta) = \prod_{m_i\in\{m_\text{sink}\}} \frac{\dd N}{\dd m}(m_i;\boldsymbol\theta).
\end{align}
In other words, $P(\{m_\text{sink}\}|\boldsymbol\theta)$ is the probability density for the particular set of sink particle masses $\{m_\text{sink}\}$ produced in our simulations, given a proposed set of parameters $\boldsymbol{\theta}$ describing the IMF. The advantage of this approach, compared to fitting a model to the histograms, is that fitting to histograms often produces results that are sensitive to the choice of bins, particularly in sparsely-populated ranges of mass; our Bayesian approach removes the need for binning.

Fitting requires some care with respect to the choice of priors. We adopt flat, uninformative priors for $m_0$, $\sigma$, and $\tan^{-1}\Gamma$, with the latter being equivalent to assuming that all angles of the power-law slope (straight line in log-log space) are equally likely \citep{Jeffreys46a}. These choices have little impact on the results of the fit parameters. For the N2 SMF, we also adopt a flat prior for $m_T$, and we obtain a good fit by doing so; we show the results of our MCMC fit in comparison to the data in Fig.~\ref{fig:imf}, indicating that the fit describes the data well. We find the high-mass slope $\Gamma(n=1.9)=-3.07\substack{+0.67 \\ -0.77}$ for N2, where the central estimate is the median of the posterior PDF, and the error bars indicate the 2nd to 98th percentile confidence interval. If we adopt a similar flat, unconstrained prior for $m_T$ for N1, we find a higher value for $m_T$ than for N2. In order to enable a meaningful comparison of the slopes between N1 and N2, we therefore adopt an informative prior on $m_T$ when fitting the N1 SMF, by setting it equal to a Gaussian approximation of the posterior distribution of $m_T$ in N2.\footnote{To be precise, the prior distribution we adopt for $m_T$ is $p_{\rm prior} \propto \exp[-(m_T-m_{T,{\rm N2,med}})^2/2\sigma_{\rm N2}^2]$, where $m_{T,{\rm N2,med}}$ is the median posterior value of $m_T$ for our fit to N2, and $\sigma_{\rm N2}$ is half the 16th to 84th percentile range for the posterior.} Intuitively, this amounts to saying that, in order to perform a meaningful comparison of slopes between N1 and N2, we demand that the turnover point $m_T$ between the lognormal and power-law portions of the SMF be at similar masses. With this prior, we find $\Gamma(n=1) = -1.20\substack{+0.23 \\ -0.27}$ for N1. We show this fit in Fig.~\ref{fig:imf}, and find that the resulting functional form is a good fit to the simulated mass distribution.

In summary, we find that the turbulence power spectrum is a key ingredient for controlling the high-mass region of the IMF, with N1 producing more massive stars than N2. The high-mass slope ($\Gamma$) of the IMF is significantly shallower for N1 compared with N2, with the Salpeter slope in between N1 and N2. We discuss possible reasons for this when we now compare the simulation results with the predictions of the IMF theories.

\section{Comparison with theoretical models of the IMF} \label{sec:models}

\begin{figure*}
    \centering
    \includegraphics[width=\linewidth]{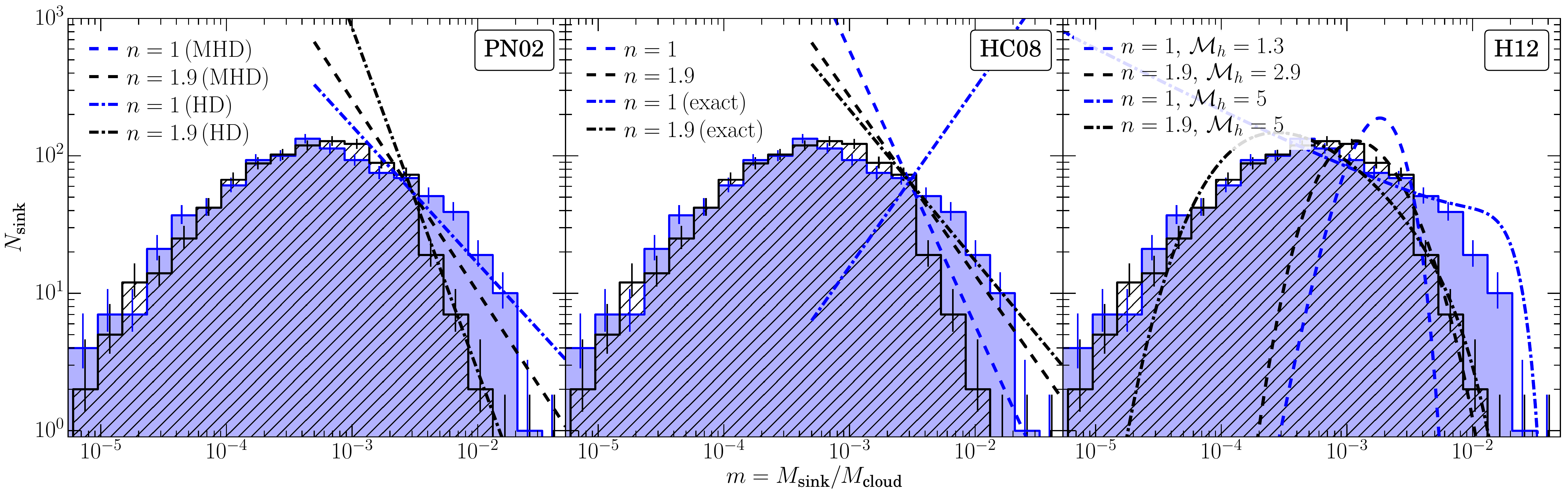}
    \caption{Comparison between the mass distributions obtained in the simulations (Fig.~\ref{fig:imf}) and the high-mass IMF slopes estimated by the three IMF theories by PN02 (left), HC08 (middle), and H12 (right). The blue histograms and lines correspond to the N1 ($n=1$) simulations and the black ones correspond to the N2 ($n=1.9$) simulations.
    Left panel: the PN02 model, using the MHD (dashed lines) or HD (dash-dotted lines) shock jump conditions. Note that the IMF slope is identical for $n=1$, regardless of the choice of the jump condition.
    Middle panel: the IMF slopes originally presented in the HC08 paper (dashed lines) and the slopes including the correction term as given in \citet[dash-dotted lines]{Hennebelle2009AnalyticalFlow}. For the PN02 and HC08 models, we anchor the power-law functions at $m=m_T=\num{3e-3}$, as obtained from the MCMC fitting of the simulation data.
    Right panel: the IMFs predicted by the H12 model, with the characteristic Mach number ($\mach_h$) calculated from the velocity dispersion on the largest scales in our simulations, $1<k<3$, where $k=1$ corresponds to the box scale $L$ (dashed lines), or set to the rms Mach number of the simulations (dash-dotted lines). We arbitrarily shift both dashed lines to lower masses by a factor of 2 and both dash-dotted lines to lower masses by a factor of 10, compared with the direct prediction of the H12 model, as an attempt to match the high-mass end of the SMFs with that of the corresponding IMFs.}
    \label{fig:theory}
\end{figure*}

In this Section we compare the simulation results with the three turbulence-regulated IMF models: PN02 \citep{Padoan2002TheFragmentation}, HC08 \citep{Hennebelle2008AnalyticalCores}, and H12 \citep{Hopkins2012TheDistribution}. We summarise the comparison in Figure~\ref{fig:theory}, as well as in Table~\ref{tab:gamma}, which lists the high-mass IMF slopes estimated from the three theoretical models and calculated from our simulations for velocity power spectral indices of $n=1$ and $1.9$. We emphasise that we only compare the high-mass region of the IMF, and other features of the IMF such as the IMF peak and the sub-stellar mass function are out of the scope of this study, since we do not include the relevant physics in our simulations (\S\ref{sec:imf}).

\begin{table}
    \caption{Predictions of the slope $\Gamma$ of the high-mass tail of the IMF from turbulence-regulated IMF theories.}
    \def\arraystretch{1.0}
    \setlength{\tabcolsep}{19.5pt}
    \begin{tabular}{rrr}
    \toprule
    model & \multicolumn{2}{r}{velocity spectral index} \\
    & $n=1$ & $n=1.9$ \\
    \midrule
    & \multicolumn{2}{l}{$\Gamma=$} \\
    PN02 & $-1.0$ & $-1.4$ \\
    PN02~(HD) & $-1.0$ & $-2.5$ \\
    HC08 & $-2.0$ & $-1.3$ \\
    HC08~(exact) & $+1.3$ & $-1.1$ \\
    H12 ($k=1-3$) & $-16$ & $-2.1$ \\
    H12 (rms) & $-0.3$ & $-2.0$ \\
    \midrule
    this study & $-1.20\substack{+0.23 \\ -0.27}$ & $-3.07\substack{+0.67 \\ -0.77}$\\
    \bottomrule
    \end{tabular} \vspace{0.1cm} \\
    \emph{Notes.} PN02: \citet{Padoan2002TheFragmentation}. PN02~(HD): PN02 with hydrodynamic shock jump conditions ($\rho'/\rho = \mach^2$). HC08: \citet{Hennebelle2008AnalyticalCores}. HC08~(exact): HC08 with the correction term discussed in \citet{Hennebelle2009AnalyticalFlow}. H12~($k=1-3$): \citet{Hopkins2012TheDistribution}, with $\mach_h$ derived by integrating the power spectrum from $k=1-3$, and slope derived by averaging between $m=\num{3e-3}$ and $10^{-2}$. H12~(rms): same as H12~($k=1-3$), but using the full rms Mach number for $\mach_h$.
    \label{tab:gamma}
\end{table}

\subsection{PN02 model} \label{sec:pn02}

In the PN02 theory, cores emerge from turbulent shocks sweeping through the molecular cloud medium, and hence the resultant IMF is dependent on the extent to which shocks compress the gas. PN02 predict that the resulting IMF will be a power law with slope
\begin{align}
    \Gamma = -3/(4-n),
\end{align}
assuming a linear shock jump condition, i.e., shocks increase the density of the gas linearly with the Mach number of the shock (hereafter ``MHD condition``). On the other hand, \citet{Padoan2007TwoFormation} suggested that if there are no magnetic fields present, it is more appropriate to consider the post-shock gas density to be proportional to $\mach^2$ (``HD condition``), which leads to 
\begin{align}
    \Gamma = -3/(5-2n).
\end{align}
In either the HD or MHD cases, PN02 predict that a shallower velocity power spectrum produces a shallower high-mass IMF: $\Gamma(n=1)=-1$ (for both the MHD and HD condition) and $\Gamma(n=1.9)=-1.4$ (MHD condition) or $-2.5$ (HD condition). We show these theoretical predictions for $\Gamma$ together with the simulation SMFs in the left-hand panel of Fig.~\ref{fig:theory}. Overall, the PN02 prediction with the HD shock jump condition (i.e. in the absence of magnetic fields) is quantitatively consistent with both N1 and N2 simulations within the 95\% interval range.

The $n$-dependence on the high-mass slope of the PN02 model comes from the linewidth-size relation. Shocks larger in size (i.e., also with higher Mach number) can sweep up more gas and thereby produce more massive cores. However, this effect is countered by the fact that shocks with higher $\mach$ produce thinner compressed post-shock layers, which reduces the mass of the resultant dense core, because the core size is set equal to the post-shock length scale in the PN02 model. Because the velocity power spectrum controls how the velocity dispersion scales with size, namely $\mach(\ell) \propto \ell^{(n-1)/2}$, altering $n$ changes the mass of cores produced by a shock with fixed length, and hence changes the IMF shape. In addition, since more massive stars take longer to form because they require a larger core with a longer dynamical time, a shallower IMF is predicted for $n=1$ in the PN02 model, which is also consistent with our finding of a lower star formation rate for $n=1$.

\subsection{HC08 model} \label{sec:hc08}

In the HC08 model, turbulence has two roles during the star formation process: it creates dense patches of gas that may become self-gravitating, but also provides additional turbulent energy that counteracts collapse. According to the model, decreasing $n$ (i.e., making the power spectrum flatter) and hence enhancing turbulence on smaller scales both narrows the density PDF (i.e., creating dense regions less frequently) and increases the critical density for collapse. This prediction suggests that the SFR would be much lower for $n=1$, consistent with our results (see \S\ref{sec:stats}, \ref{sec:sfr}). HC08 also predict\footnote{Here we note that our $n$ is the index of the one-dimensional power spectrum, whereas HC08 work in terms of the three-dimensional spectrum, which has index $n-2$. Care should therefore be taken in comparing the expressions we give here to those given in HC08, since our $n$ does not refer to the same quantity as the $n$ that appears in their equations.}
\begin{align}
    \Gamma \approx -(n+3)/(2n), \label{eq:hc08}
\end{align}
that is, turbulence with a shallower velocity power spectrum produces a steeper IMF, which is opposite to what is observed in our simulations (middle panel of Fig.~\ref{fig:theory}).

However, \citet{Hennebelle2009AnalyticalFlow} suggested a correction term for Eq.~(\ref{eq:hc08}):
\begin{align}
    \Gamma = -\frac{n+3}{2n} + \frac{3(3-n)}{n}\frac{\ln{\mach_*}}{\sigma_s^2}, \label{eq:hc09}
\end{align}
where $\mach_*$ is the (one-dimensional) Mach number on the Jeans scale ($\lambda_J$) and $\sigma_s^2$ is the global variance in the logarithmic density $s$. Under usual circumstances, where $n\approx 2$ and $\mach_*\lesssim 10$, the second term is close to zero and has only minimal effect on the overall shape of the IMF. However, for $n=1$, the correction term becomes much more significant. We calculate the exact value of the high-mass slope predicted by HC08 with the correction term to be $\Gamma=+1.3$ for $n=1$, given in our $n=1$ simulations $\sigma_s^2 = 1.94$ at the beginning of gravitational collapse ($t=2\,T$) and $\mach_* = 4.9 / 3^{1/2} = 2.8$ (converting the 3D Mach number of $\sim4.9$ in the simulations, to the 1D Mach number used in the HC model). While the correction is in the right direction, it is far larger than the difference between the measured value from our simulations and the HC08 prediction, and appears implausible, since for $\Gamma=1.3$ the total mass in the high-mass tail of the IMF would diverge.

\subsection{H12 model} \label{sec:h12}

The role of the velocity power spectrum in the H12 theory is similar to that in the HC08 theory. The primary difference between the theories lies in how one estimates the density PDF and counts the number of bound regions as a function of length scale. The difference is nonetheless significant; for example, H12 speculates that the density variance is greater on small length scales and smaller on large length scales for $n\approx 1$, qualitatively similar to our results (Fig.~\ref{fig:psdens}), while in HC08 the density variance is smaller across all scales. Since the H12 IMF model generally does not have a closed form, one needs to follow the excursion-set formalism and directly rebuild the mass functions in order to study the effect of $n$ in the H12 model. We therefore developed our own Python code that reproduces the last-crossing IMF, and compared the results with our simulation.\footnote{We make one modification in our code relative to the original H12 model. In the H12 model, the barrier function includes a term representing rotational support, parameterised by the epicyclic frequency $\kappa$. Since our simulation has no systematic rotation, we take the limit $\kappa\to 0$ when evaluating the barrier function.}

In the H12 theory, the power spectral index $n$ and the characteristic Mach number $\mach_h$ are the two important parameters that determine the shape of the IMF. The parameter $n$ is straightforward to define and measure in our simulations, but there is some ambiguity in how to define $\mach_h$ for our simulation. In the context of the H12 model, $\mach_h$ is the Mach number of the velocity field measured on sizes comparable to the galactic scale height, $h$, which is identified with the outer scale of the turbulent cascade. Our simulation does not possess a scale height, since it takes place in a periodic box, and there is some ambiguity in how to define the outer scale of the turbulence, particularly for the $n=1$ case where turbulent power is not sharply peaked on large scales. We therefore consider two possibilities, which roughly bracket the range of reasonable choices. The first is simply to set $\mach_h = \mach = 5$, i.e., to set the Mach number at the outer scale of the turbulence equal to the Mach number of the simulation box as a whole. This choice is most consistent with the implicit assumption in the H12 model that the turbulent power is mostly on large scales, so as one considers larger and larger size scale, the Mach number monotonically increases, approaching the total Mach number as the size scale under consideration approaches $h$. Our second method for estimating $\mach_h$ is to integrate the velocity power spectra in the region $1<k<3$, which is roughly the outer scale of the turbulence in our periodic box. Doing so, we find $\mach_h=1.3$ for the N1 simulations and $2.9$ for the N2 simulations. 

We compare the predictions of the H12 model with the aforementioned parameters to our simulations in the right-hand panel of Fig.~\ref{fig:theory} (dashed and dot-dashed lines). We first focus on the case where we measure $\mach_h$ by integrating over $k=1-3$, and observe that, while the IMF predicted for $n=1.9$ coincides fairly well with the N2 simulations for $m\gtrsim 10^{-3}$, the $n=1$ prediction is significantly steeper than that for $n=1.9$, which is the opposite of what we observe from our simulations. By contrast, if we accept a mass shifting factor\footnote{A possible justification for this shift is that in our simulations there are no density fluctuations at the box scale, whereas in the H12 model fluctuations at the galactic scale height $h$ are non-zero, and only damp to zero on scales $\ll h$ \citep[e.g.][Fig.~2]{Hopkins2013AFragmentation}.} of $10$, the predicted IMF shapes beyond the peaks are significantly closer to what we measure for both the N1 and N2 simulations in the case where we take $\mach_h=\mach=5$ (dash-dotted lines), except near $m\approx 10^{-2}$. The predicted qualitative effect of varying $n$ is also consistent with our simulation results, and with \citet{Hopkins2013AFragmentation}. According to the H12 model, the cutoff in the N1 SMF beyond $m>10^{-2}$, which is most likely a result of the finite mass in the simulation box, is explained by the suppression of density fluctuations due to mass conservation. However, we caution that, because of the ambiguity in the definition of $\mach_h$ inherent in the H12 models, as well as the necessity of an arbitrary horizontal shift, we can only tentatively identify this as a successful prediction. Finally, we note that while the H12 model in principle allows for the inclusion of magnetic fields, the dependence of the IMF on the magnetic field has not been studied in detail in \citet{Hopkins2013AFragmentation}. We aim to quantify the effects of the magnetic field on the IMF in a follow-up study.

\section{Conclusions} \label{sec:sum}

Using hydrodynamical simulations that include gravity and sink particles, we investigate the effect of the shape of the power spectrum of supersonic turbulence ($E_v(k)\propto k^{-n}$) on the stellar IMF. With the help of adaptive mesh refinement and repeated simulations with different random seeds for the turbulence, we construct statistically significant sink mass distributions with 900--1000 sink particles formed for each $n$, and a dynamic range spanning three orders of magnitude, from a low-mass cutoff imposed by the grid resolution to a high-mass cutoff imposed by the finite size of the simulation domain. From the sink particle populations, we find that turbulence with $n=1$ significantly flattens the high-mass end of the IMF compared to $n\approx 2$ (i.e., $n=1$ turbulence generates more massive stars), but has little effect on the distribution of low-mass stars and sub-stellar objects. This result is consistent with our current understanding of molecular cloud dynamics and star formation: turbulence governs the large-scale fragmentation of molecular clouds, while other mechanisms such as radiative heating play more important roles below a certain length (or mass) scale. We also find that compared to natural supersonic turbulence with $n\approx 2$, turbulence with a scaling index of $n=1$ creates less density dispersion, does not promote the formation of large-scale gas structures such as large-scale filaments, and slows down the star formation rate.

We compare our simulation results with three turbulence-regulated theoretical models of the IMF: \citet[PN02]{Padoan2002TheFragmentation}, \citet[HC08]{Hennebelle2008AnalyticalCores}, and \citet[H12]{Hopkins2012TheDistribution}. We find that the qualitative predictions of the three models vary significantly (e.g., the dependence of the high-mass slope of the IMF on $n$). Out of the three IMF models, we find that the PN02 theory is consistent with our measurement of the $n$-dependence of the high-mass IMF slope ($\Gamma$). The density statistics predicted by the HC08 model agree qualitatively with our observations, but their predicted high-mass slope diverges for $n\to 1$. We find that the H12 model can be made similar to our simulated IMFs in the high-mass range. However, the model is quite sensitive to the choice of the definition of a key parameter ($\mach_h$), which is defined somewhat ambiguously in the model, and if we adopt an alternative definition, the H12 theory predicts qualitatively different results that disagree with our simulations.

There remains one important question that is not yet answered: why did turbulence with $n\approx 2$ shape a high-mass IMF much steeper than the Salpeter IMF in our simulations? As mentioned in \S\ref{sec:intro} and \S\ref{sec:pn02}, the answer may be the absence of magnetic fields, since only the PN02 theory successfully predicts the high-mass slope for the $n\approx 2$ hydrodynamical turbulence (apart from the modified H12 theory with $\mach_h=5$), and it is the only model that explicitly encodes the role of magnetic fields in shaping the high-mass IMF. We suggest a follow-up study that includes varying levels of magnetic fields, in order to quantify the role of the magnetic field on the shape of the IMF.

\section*{Acknowledgements}
We thank {\AA}ke Nordlund for providing a detailed and constructive referee report. We also thank Patrick Hennebelle and Paolo Padoan for their interest, comments and suggestions on the manuscript. We further thank Phil Hopkins and D{\'a}vid Guszejnov for their help with reproducing the H12 IMF model. C.~F.~acknowledges funding provided by the Australian Research Council (Discovery Project DP170100603 and Future Fellowship FT180100495), and the Australia-Germany Joint Research Cooperation Scheme (UA-DAAD). M.~R.~K.~acknowledges funding from the Australian Research Council (Discovery Project DP190101258 and Future Fellowship FT180100375), and the Australia-Germany Joint Research Cooperation Scheme (UA-DAAD). We further acknowledge high-performance computing resources provided by the Leibniz Rechenzentrum and the Gauss Centre for Supercomputing (grants~pr32lo, pr48pi and GCS Large-scale project~10391), the Australian National Computational Infrastructure (grants~ek9 and jh2) in the framework of the National Computational Merit Allocation Scheme and the ANU Merit Allocation Scheme. The simulation software FLASH was in part developed by the DOE-supported Flash Center for Computational Science at the University of Chicago.

\section*{Data Availability}

The simulation data underlying this article will be shared on reasonable request to Donghee Nam at \href{mailto:u6836819@anu.edu.au}{u6836819@anu.edu.au}. Our Python code that reproduces the H12 last-crossing IMF is publicly available at \href{https://github.com/dongheenam/hopkins-imf}{https://github.com/dongheenam/hopkins-imf}.




\bibliographystyle{mnras}
\bibliography{bibliography} 

\begin{thebibliography}{}
\makeatletter
\relax
\def\mn@urlcharsother{\let\do\@makeother \do\$\do\&\do\#\do\^\do\_\do\%\do\~}
\def\mn@doi{\begingroup\mn@urlcharsother \@ifnextchar [ {\mn@doi@}
  {\mn@doi@[]}}
\def\mn@doi@[#1]#2{\def\@tempa{#1}\ifx\@tempa\@empty \href
  {http://dx.doi.org/#2} {doi:#2}\else \href {http://dx.doi.org/#2} {#1}\fi
  \endgroup}
\def\mn@eprint#1#2{\mn@eprint@#1:#2::\@nil}
\def\mn@eprint@arXiv#1{\href {http://arxiv.org/abs/#1} {{\tt arXiv:#1}}}
\def\mn@eprint@dblp#1{\href {http://dblp.uni-trier.de/rec/bibtex/#1.xml}
  {dblp:#1}}
\def\mn@eprint@#1:#2:#3:#4\@nil{\def\@tempa {#1}\def\@tempb {#2}\def\@tempc
  {#3}\ifx \@tempc \@empty \let \@tempc \@tempb \let \@tempb \@tempa \fi \ifx
  \@tempb \@empty \def\@tempb {arXiv}\fi \@ifundefined
  {mn@eprint@\@tempb}{\@tempb:\@tempc}{\expandafter \expandafter \csname
  mn@eprint@\@tempb\endcsname \expandafter{\@tempc}}}

\bibitem[\protect\citeauthoryear{Andr{\'{e}} et~al.,}{Andr{\'{e}}
  et~al.}{2010}]{Andre2010FromSurvey}
Andr{\'{e}} P.,  et~al., 2010, A{\&}A, 518, L102

\bibitem[\protect\citeauthoryear{Bastian, Covey  \& Meyer}{Bastian
  et~al.}{2010}]{Bastian2010AVariations}
Bastian N.,  Covey K.~R.,   Meyer M.~R.,  2010, \mn@doi [ARA{\&}A]
  {10.1146/annurev-astro-082708-101642}, 48, 339

\bibitem[\protect\citeauthoryear{Bate}{Bate}{2009}]{Bate2009TheStructure}
Bate M.~R.,  2009, \mn@doi [MNRAS] {10.1111/j.1365-2966.2009.14970.x}, 397, 232

\bibitem[\protect\citeauthoryear{Bertoldi \& McKee}{Bertoldi \&
  McKee}{1992}]{Bertoldi1992Pressure-confinedClouds}
Bertoldi F.,  McKee C.~F.,  1992, \mn@doi [J. Chem. Inf. Model.]
  {10.1017/CBO9781107415324.004}, 395, 140

\bibitem[\protect\citeauthoryear{Bond, Cole, Efstathiou  \& Kaiser}{Bond
  et~al.}{1991}]{Bond1991ExcursionFluctuations}
Bond J.~R.,  Cole S.,  Efstathiou G.,   Kaiser N.,  1991, ApJ, p.~440

\bibitem[\protect\citeauthoryear{Bouchut, Klingenberg  \& Waagan}{Bouchut
  et~al.}{2010}]{Bouchut2010AWaves}
Bouchut F.,  Klingenberg C.,   Waagan K.,  2010, \mn@doi [Numer. Math.]
  {10.1007/s00211-010-0289-4}, 115, 647

\bibitem[\protect\citeauthoryear{Brunt, Heyer  \& {Mac Low}}{Brunt
  et~al.}{2009}]{Brunt2009TurbulentClouds}
Brunt C.~M.,  Heyer M.~H.,   {Mac Low} M.-M.,  2009, \mn@doi [A{\&}A]
  {10.1051/0004-6361/200911797}, 504, 883

\bibitem[\protect\citeauthoryear{Chabrier}{Chabrier}{2003}]{Chabrier2003GalacticFunction}
Chabrier G.,  2003, \mn@doi [PASP] {10.1086/376392}, 115, 763

\bibitem[\protect\citeauthoryear{Chabrier}{Chabrier}{2005}]{Chabrier2005The2005}
Chabrier G.,  2005, \mn@doi [Astrophys. Sp. Sci. Libr.]
  {10.1007/978-1-4020-3407-7_5}, 327, 41

\bibitem[\protect\citeauthoryear{Delgado-Donate, Clarke  \&
  Bate}{Delgado-Donate et~al.}{2004}]{Delgado-Donate2004TheFormation}
Delgado-Donate E.~J.,  Clarke C.~J.,   Bate M.~R.,  2004, \mn@doi [MNRAS]
  {10.1111/j.1365-2966.2004.07259.x}, 347, 759

\bibitem[\protect\citeauthoryear{Elmegreen \& Scalo}{Elmegreen \&
  Scalo}{2004}]{Elmegreen2004InterstellarProcesses}
Elmegreen B.~G.,  Scalo J.,  2004, \mn@doi [ARA{\&}A]
  {10.1146/annurev.astro.41.011802.094859}, 42, 211

\bibitem[\protect\citeauthoryear{Eswaran \& Pope}{Eswaran \&
  Pope}{1988}]{Eswaran1988AnTurbulence}
Eswaran V.,  Pope S.~B.,  1988, Comput. Fluids, 16, 257

\bibitem[\protect\citeauthoryear{Falgarone, Puget  \& P{\'{e}}rault}{Falgarone
  et~al.}{1992}]{Falgarone1992TheClouds}
Falgarone E.,  Puget J.-L.,   P{\'{e}}rault M.,  1992, A{\&}A, 257, 715

\bibitem[\protect\citeauthoryear{Federrath}{Federrath}{2013}]{Federrath2013OnTurbulence}
Federrath C.,  2013, \mn@doi [MNRAS] {10.1093/mnras/stt1644}, 436, 1245

\bibitem[\protect\citeauthoryear{Federrath}{Federrath}{2015}]{Federrath2015InefficientFeedback}
Federrath C.,  2015, \mn@doi [MNRAS] {10.1093/mnras/stv941}, 450, 4035

\bibitem[\protect\citeauthoryear{Federrath \& Klessen}{Federrath \&
  Klessen}{2012}]{Federrath2012TheObservations}
Federrath C.,  Klessen R.~S.,  2012, \mn@doi [ApJ]
  {10.1088/0004-637X/761/2/156}, 761, 156

\bibitem[\protect\citeauthoryear{Federrath, Roman-Duval, Klessen, Schmidt  \&
  {Mac Low}}{Federrath et~al.}{2010a}]{Federrath2010ComparingForcing}
Federrath C.,  Roman-Duval J.,  Klessen R.,  Schmidt W.,   {Mac Low} M.~M.,
  2010a, \mn@doi [A{\&}A] {10.1051/0004-6361/200912437}, 512

\bibitem[\protect\citeauthoryear{Federrath, Banerjee, Clark  \&
  Klessen}{Federrath et~al.}{2010b}]{Federrath2010ModelingSPH}
Federrath C.,  Banerjee R.,  Clark P.~C.,   Klessen R.~S.,  2010b, \mn@doi
  [ApJ] {10.1088/0004-637X/713/1/269}, 713, 269

\bibitem[\protect\citeauthoryear{Federrath, Sur, Schleicher, Banerjee  \&
  Klessen}{Federrath et~al.}{2011}]{Federrath2011ATurbulence}
Federrath C.,  Sur S.,  Schleicher D.~R.,  Banerjee R.,   Klessen R.~S.,  2011,
  \mn@doi [ApJ] {10.1088/0004-637X/731/1/62}, 731, 62

\bibitem[\protect\citeauthoryear{Federrath, Krumholz  \& Hopkins}{Federrath
  et~al.}{2017}]{Federrath2017ConvergingStars}
Federrath C.,  Krumholz M.,   Hopkins P.~F.,  2017, J. Phys. Conf. Ser., 837,
  012007

\bibitem[\protect\citeauthoryear{Federrath, Klessen, Iapichino  \&
  Beattie}{Federrath et~al.}{2020}]{Federrath2020TheSimulation}
Federrath C.,  Klessen R.~S.,  Iapichino L.,   Beattie J.~R.,  2020, {The sonic
  scale revealed by the world's largest super- sonic turbulence simulation}
  (\mn@eprint {arXiv} {arXiv:2011.06238v1})

\bibitem[\protect\citeauthoryear{Foreman-Mackey, Hogg, Lang  \&
  Goodman}{Foreman-Mackey et~al.}{2013}]{Foreman-Mackey2013EmceeHammer}
Foreman-Mackey D.,  Hogg D.~W.,  Lang D.,   Goodman J.,  2013, \mn@doi [PASP]
  {10.1086/670067}, 125, 306

\bibitem[\protect\citeauthoryear{Fryxell et~al.,}{Fryxell
  et~al.}{2000}]{Fryxell2000FLASHFlashes}
Fryxell B.,  et~al., 2000, Astrophys. J. Suppl. Ser., 131, 273

\bibitem[\protect\citeauthoryear{Goodwin, Whitworth  \& Ward-Thompson}{Goodwin
  et~al.}{2006}]{Goodwin2006StarSpectrum}
Goodwin S.~P.,  Whitworth A.~P.,   Ward-Thompson P.,  2006, \mn@doi [A{\&}A]
  {10.1051/0004-6361:20054026}, 452, 487

\bibitem[\protect\citeauthoryear{Guszejnov \& Hopkins}{Guszejnov \&
  Hopkins}{2015}]{Guszejnov2015MappingFunction}
Guszejnov D.,  Hopkins P.~F.,  2015, \mn@doi [MNRAS] {10.1093/mnras/stv872},
  450, 4137

\bibitem[\protect\citeauthoryear{Haugb{\o}lle, Padoan  \&
  Nordlund}{Haugb{\o}lle et~al.}{2018}]{Haugbolle2018TheTurbulence}
Haugb{\o}lle T.,  Padoan P.,   Nordlund {\AA}.,  2018, \mn@doi [ApJ]
  {10.3847/1538-4357/aaa432}, 854, 35

\bibitem[\protect\citeauthoryear{Hennebelle \& Chabrier}{Hennebelle \&
  Chabrier}{2008}]{Hennebelle2008AnalyticalCores}
Hennebelle P.,  Chabrier G.,  2008, \mn@doi [ApJ] {10.1086/589916}, 684, 395

\bibitem[\protect\citeauthoryear{Hennebelle \& Chabrier}{Hennebelle \&
  Chabrier}{2009}]{Hennebelle2009AnalyticalFlow}
Hennebelle P.,  Chabrier G.,  2009, \mn@doi [ApJ]
  {10.1088/0004-637X/702/2/1428}, 702, 1428

\bibitem[\protect\citeauthoryear{Heyer \& Brunt}{Heyer \&
  Brunt}{2004}]{Heyer2004TheClouds}
Heyer M.~H.,  Brunt C.~M.,  2004, ApJ, 615, L45

\bibitem[\protect\citeauthoryear{Hopkins}{Hopkins}{2012}]{Hopkins2012TheDistribution}
Hopkins P.~F.,  2012, \mn@doi [MNRAS] {10.1111/j.1365-2966.2012.20731.x}, 423,
  2037

\bibitem[\protect\citeauthoryear{Hopkins}{Hopkins}{2013}]{Hopkins2013AFragmentation}
Hopkins P.~F.,  2013, \mn@doi [MNRAS] {10.1093/mnras/sts704}, 430, 1653

\bibitem[\protect\citeauthoryear{Hopkins}{Hopkins}{2018}]{Hopkins2018TheFunction}
Hopkins A.~M.,  2018, \mn@doi [Publ. Astron. Soc. Aust.]
  {10.1017/pas.2018.xxx}, 35

\bibitem[\protect\citeauthoryear{Jeffreys}{Jeffreys}{1946}]{Jeffreys46a}
Jeffreys H.,  1946, \mn@doi [Proc. R. Soc. London. Series A. Math. Phys. Sci.]
  {10.1098/rspa.1946.0056}, 186, 453

\bibitem[\protect\citeauthoryear{Kim \& Ryu}{Kim \&
  Ryu}{2005}]{Kim2005DensityFlows}
Kim J.,  Ryu D.,  2005, \mn@doi [ApJ] {10.1086/491600}, 630, L45

\bibitem[\protect\citeauthoryear{Konstandin, Schmidt, Girichidis, Peters,
  Shetty  \& Klessen}{Konstandin et~al.}{2016}]{Konstandin2016MachField}
Konstandin L.,  Schmidt W.,  Girichidis P.,  Peters T.,  Shetty R.,   Klessen
  R.~S.,  2016, \mn@doi [MNRAS] {10.1093/mnras/stw1313}, 460, 4483

\bibitem[\protect\citeauthoryear{Kritsuk, Norman, Padoan  \& Wagner}{Kritsuk
  et~al.}{2007}]{Kritsuk2007TheTurbulence}
Kritsuk A.~G.,  Norman M.~L.,  Padoan P.,   Wagner R.,  2007, \mn@doi [ApJ]
  {10.1086/519443}, 665, 416

\bibitem[\protect\citeauthoryear{Kroupa}{Kroupa}{2001}]{Kroupa2001OnFunction}
Kroupa P.,  2001, MNRAS, 322, 231

\bibitem[\protect\citeauthoryear{Kroupa, Weidner, Pflamm-Altenburg, Thies,
  Dabringhausen, Marks  \& Maschberger}{Kroupa
  et~al.}{2013}]{Kroupa2013ThePopulations}
Kroupa P.,  Weidner C.,  Pflamm-Altenburg J.,  Thies I.,  Dabringhausen J.,
  Marks M.,   Maschberger T.,  2013, in , Vol.~5, Oswalt T.D., Gilmore G.
  Planets, Stars Stellar Syst. Springer, Dordrecht..
Springer Science+Business Media Dordrecht, p.~115 (\mn@eprint {arXiv}
  {1112.3340}), \mn@doi{10.1007/978-94-007-5612-0_4}

\bibitem[\protect\citeauthoryear{Krumholz}{Krumholz}{2011}]{Krumholz2011OnMasses}
Krumholz M.~R.,  2011, \mn@doi [ApJ] {10.1088/0004-637X/743/2/110}, 743

\bibitem[\protect\citeauthoryear{Krumholz}{Krumholz}{2014}]{Krumholz2014TheFunction}
Krumholz M.~R.,  2014, \mn@doi [Phys. Rep.] {10.1016/j.physrep.2014.02.001},
  539, 49

\bibitem[\protect\citeauthoryear{Krumholz \& Federrath}{Krumholz \&
  Federrath}{2019}]{Krumholz2019TheFunction}
Krumholz M.~R.,  Federrath C.,  2019, Frontiers in Astronomy and Space
  Sciences, 6, 7

\bibitem[\protect\citeauthoryear{Krumholz, McKee  \& Klein}{Krumholz
  et~al.}{2004}]{Krumholz2004EmbeddingGrids}
Krumholz M.~R.,  McKee C.~F.,   Klein R.~I.,  2004, \mn@doi [ApJ]
  {10.1086/421935}, 611, 399

\bibitem[\protect\citeauthoryear{Krumholz, Myers, Klein  \& McKee}{Krumholz
  et~al.}{2016}]{Krumholz2016WhatSimulations}
Krumholz M.~R.,  Myers A.~T.,  Klein R.~I.,   McKee C.~F.,  2016, \mn@doi
  [MNRAS] {10.1093/mnras/stw1236}, 460

\bibitem[\protect\citeauthoryear{Larson}{Larson}{1981}]{Larson1981TurbulenceClouds}
Larson R.~B.,  1981, MNRAS, 194, 809

\bibitem[\protect\citeauthoryear{{Mac Low}, Smith, Klessen  \& Burkert}{{Mac
  Low} et~al.}{1998}]{MacLow1998TheClouds}
{Mac Low} M.,  Smith M.~D.,  Klessen R.~S.,   Burkert A.,  1998, Ap{\&}SS, 261,
  195

\bibitem[\protect\citeauthoryear{Mathew \& Federrath}{Mathew \&
  Federrath}{2020}]{Mathew2020ImplementationFunction}
Mathew S.~S.,  Federrath C.,  2020, \mn@doi [MNRAS] {10.1093/mnras/staa1931},
  496, 5201

\bibitem[\protect\citeauthoryear{McKee \& Ostriker}{McKee \&
  Ostriker}{2007}]{McKee2007TheoryFormation}
McKee C.~F.,  Ostriker E.~C.,  2007, \mn@doi [ARA{\&}A]
  {10.1146/annurev.astro.45.051806.110602}, 45, 565

\bibitem[\protect\citeauthoryear{Miller \& Scalo}{Miller \&
  Scalo}{1979}]{Miller1979TheNeighborhood}
Miller G.~E.,  Scalo J.~M.,  1979, Astrophys. J. Suppl. Ser., 41, 513

\bibitem[\protect\citeauthoryear{Offner, Klein, McKee  \& Krumholz}{Offner
  et~al.}{2009}]{Offner2009TheFormation}
Offner S.~S.,  Klein R.~I.,  McKee C.~F.,   Krumholz M.~R.,  2009, \mn@doi
  [ApJ] {10.1088/0004-637X/703/1/131}, 703, 131

\bibitem[\protect\citeauthoryear{Offner, Clark, Hennebelle, Bastian, Bate,
  Hopkins, Moraux  \& Whitworth}{Offner et~al.}{2014}]{Offner2014TheFunction}
Offner S. S.~R.,  Clark P.~C.,  Hennebelle P.,  Bastian N.,  Bate M.~R.,
  Hopkins P.~F.,  Moraux E.,   Whitworth A.~P.,  2014, \mn@doi [Protostars
  Planets VI] {10.2458/azu_uapress_9780816531240-ch003}, 914, 53

\bibitem[\protect\citeauthoryear{Ossenkopf \& {Mac Low}}{Ossenkopf \& {Mac
  Low}}{2002}]{Ossenkopf2002TurbulentClouds}
Ossenkopf V.,  {Mac Low} M.-M.,  2002, \mn@doi [A{\&}A]
  {10.1051/0004-6361:20020629}, 390, 307

\bibitem[\protect\citeauthoryear{Padoan \& Nordlund}{Padoan \&
  Nordlund}{2002}]{Padoan2002TheFragmentation}
Padoan P.,  Nordlund {\AA}.,  2002, \mn@doi [ApJ] {10.1086/341790}, 576, 870

\bibitem[\protect\citeauthoryear{Padoan, Nordlund  \& Jones}{Padoan
  et~al.}{1997}]{Padoan1997TheFunction}
Padoan P.,  Nordlund {\AA}.,   Jones B. J.~T.,  1997, MNRAS, 288, 145

\bibitem[\protect\citeauthoryear{Padoan, Nordlund, Kritsuk, Norman  \&
  Li}{Padoan et~al.}{2007}]{Padoan2007TwoFormation}
Padoan P.,  Nordlund {\AA}.,  Kritsuk A.~G.,  Norman M.~L.,   Li P.~S.,  2007,
  \mn@doi [ApJ] {10.1086/516623}, 661, 972

\bibitem[\protect\citeauthoryear{Padoan, Haugbølle  \& Nordlund}{Padoan
  et~al.}{2014}]{Padoan2014Infall-drivenProblem}
Padoan P.,  Haugbølle T.,   Nordlund {\AA}.,  2014, \mn@doi [ApJ]
  {10.1088/0004-637X/797/1/32}, 797, 32

\bibitem[\protect\citeauthoryear{Press \& Schechter}{Press \&
  Schechter}{1974}]{Press1974FormationCondensation}
Press W.~H.,  Schechter P.,  1974, ApJ, 187, 425

\bibitem[\protect\citeauthoryear{Ricker}{Ricker}{2008}]{Ricker2008AMeshes}
Ricker P.~M.,  2008, \mn@doi [Astrophys. J. Suppl. Ser.] {10.1086/526425}, 176,
  293

\bibitem[\protect\citeauthoryear{Roman-Duval, Federrath, Brunt, Heyer, Jackson
  \& Klessen}{Roman-Duval et~al.}{2011}]{Roman-Duval2011TheCalibration}
Roman-Duval J.,  Federrath C.,  Brunt C.,  Heyer M.,  Jackson J.,   Klessen
  R.~S.,  2011, \mn@doi [ApJ] {10.1088/0004-637X/740/2/120}, 740, 120

\bibitem[\protect\citeauthoryear{Salpeter}{Salpeter}{1955}]{Salpeter1955TheEvolution}
Salpeter E.~E.,  1955, ApJ, p.~161

\bibitem[\protect\citeauthoryear{Stone, Ostriker  \& Gammie}{Stone
  et~al.}{1998}]{Stone1998DissipationTurbulence}
Stone J.~M.,  Ostriker E.~C.,   Gammie C.~F.,  1998, \mn@doi [ApJ]
  {10.1086/311718}, 508, L99

\bibitem[\protect\citeauthoryear{Truelove, Klein, McKee, {Holliman II}, Howell
  \& Greenough}{Truelove et~al.}{1997}]{Truelove1997TheHydrodynamics}
Truelove J.~K.,  Klein R.~I.,  McKee C.~F.,  {Holliman II} J.~H.,  Howell
  L.~H.,   Greenough J.~A.,  1997, \mn@doi [ApJ] {10.1086/310975}, 489, L179

\bibitem[\protect\citeauthoryear{Waagan, Federrath  \& Klingenberg}{Waagan
  et~al.}{2011}]{Waagan2011ATests}
Waagan K.,  Federrath C.,   Klingenberg C.,  2011, \mn@doi [J. Comput. Phys.]
  {10.1016/j.jcp.2011.01.026}, 230, 3331

\makeatother
\end{thebibliography}




\appendix

\section{Effect of turbulent driving range on the IMF shape}
\label{app:driverange}

\begin{figure}
    \centering
    \includegraphics[width=\columnwidth]{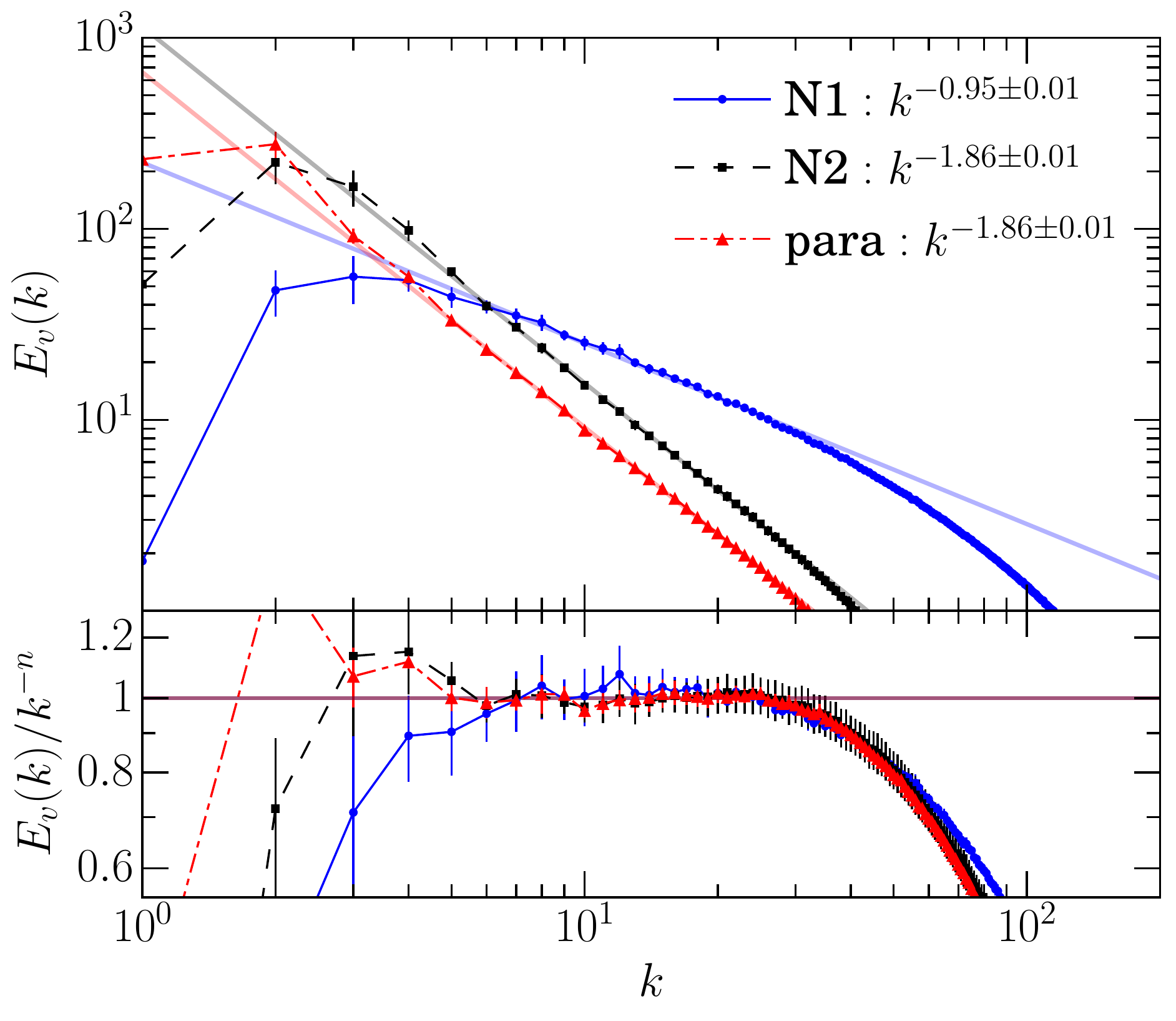}
    \caption{Turbulent velocity power spectra $E_v(k)$ (top) and the compensated power spectra $E_v(k)/k^{-n}$ of the N1 (blue solid line), N2 (black dashed line) and para (red dash-dotted line) simulations. The errorbars and fitting methods are identical to those used in Fig.~\ref{fig:ps}.}
    \label{fig:ps_para}
\end{figure}

\begin{figure}
    \centering
    \includegraphics[width=\columnwidth]{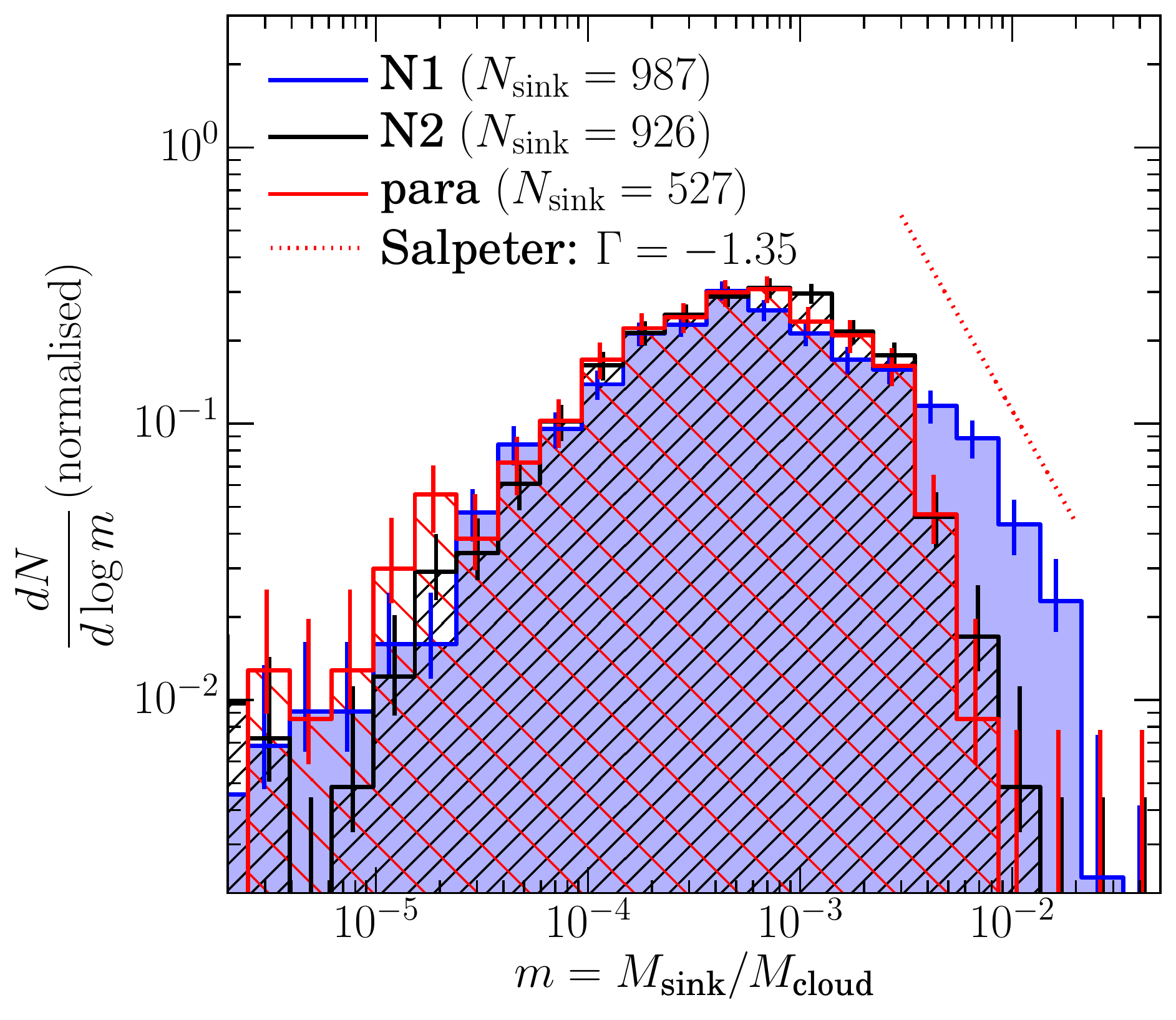}
    \caption{Logarithmic mass function $\dd N/ \dd\log{m}$ of the sink particles from the N1 (blue shaded histogram) and N2 (black densely-hatched histogram) simulations as in the main part of the article, and with the \emph{para} (red coarsely-hatched histogram) simulations added (all at $\text{SFE}=10\%$). To facilitate the comparison, we normalise the histograms so that the total area under the histogram is $1$ for all three cases. The figure is otherwise identical to Fig.~\ref{fig:imf}.}
    \label{fig:imf_para}
\end{figure}

Here we compare the N2 simulations, in which the turbulence is driven with $n\approx 2$ over an extended wavenumber range of $2\le k \le 256$, with an additional set of simulations in which only large-scale modes ($1<k<3$) are excited and the turbulent cascade naturally populates the small-scale modes \citep[i.e., as in][for example]{Federrath2015InefficientFeedback,Mathew2020ImplementationFunction}. We run four simulations with this large-scale driving (hereafter denoted as \emph{para} simulations) to check the effect of our turbulence driving method on the velocity power spectra and the IMF.

In Fig.~\ref{fig:ps_para} we show the velocity power spectra of the N1, N2 and \emph{para} simulations. We measure the scaling exponent of the velocity power spectrum in the \emph{para} simulations and find $E_v(k) \propto k^{\nume{-1.86(1)}}$, which is identical to that in the N2 simulations. The \emph{para} simulations have more power in very large modes ($k=1-2$) compared to the N2 simulations because most of the energy is injected on those scales. In Fig.~\ref{fig:imf_para} we compare the SMFs from the three simulation sets. We find that the SMFs from the N2 and \emph{para} simulations are statistically indistinguishable. Therefore, we conclude that the choice of the turbulent driving range for $n\approx 2$ does not affect the mass distribution of sink particles formed in simulations with $n\approx 2$, as this is the turbulence exponent that naturally arises when driving supersonic turbulence on large scales \citep{Federrath2013OnTurbulence,Federrath2020TheSimulation}.


\bsp	
\label{lastpage}
\end{document}